\documentclass{article}

\usepackage[english]{babel}

\usepackage[a4paper,top=2cm,bottom=2cm,left=3cm,right=3cm,marginparwidth=1.75cm]{geometry}

\usepackage{amsmath}
\usepackage[round]{natbib}
\usepackage{graphicx}
\usepackage{censor}
\StopCensoring
\usepackage[colorlinks=true, allcolors=blue]{hyperref}
\usepackage{booktabs}
\usepackage{multirow}
\usepackage[flushleft]{threeparttable}
\usepackage{comment}

\title{Detecting The Corruption Of Online Questionnaires By Artificial Intelligence}
\author{Benjamin Lebrun, Sharon Temtsin, Andrew Vonasch, Christoph Bartneck}

\begin{document}
\maketitle

\begin{abstract}
Online questionnaires that use crowd-sourcing platforms to recruit participants have become commonplace, due to their ease of use and low costs. Artificial Intelligence (AI) based Large Language Models (LLM) have made it easy for bad actors to automatically fill in online forms, including generating meaningful text for open-ended tasks. These technological advances threaten the data quality for studies that use online questionnaires. This study tested if text generated by an AI for the purpose of an online study can be detected by both humans and automatic AI detection systems. While humans were able to correctly identify authorship of text above chance level (76 percent accuracy), their performance was still below what would be required to ensure satisfactory data quality. Researchers currently have to rely on the disinterest of bad actors to successfully use open-ended responses as a useful tool for ensuring data quality. Automatic AI detection systems are currently completely unusable. If AIs become too prevalent in submitting responses then the costs associated with detecting fraudulent submissions will outweigh the benefits of online questionnaires. Individual attention checks will no longer be a sufficient tool to ensure good data quality. This problem can only be systematically addressed by crowd-sourcing platforms. They cannot rely on automatic AI detection systems and it is unclear how they can ensure data quality for their paying clients. \end{abstract}
\begin{small}
\end{small}
\section{Introduction}

Using crowd-sourcing platforms for recruiting participants for online questionnaires has always been susceptible to abuse. Bad actors could randomly click answers to quickly earn money, even at scale. Until recently, a solution to this problem was to ask online participants to complete open-ended responses that could not be completed through random button clicking. However, the development of Large Language Models, such as ChatGPT or Bard, threatens the viability of this solution. This threat to the data collection quality has to be understood in the wider context of methodological challenges that all add up to what is now famously coined the ``replication crisis''.

The replication crisis, initially observed in the field of psychology and human behavior, has also been shown to occur in other domains, including computer science, chemistry, biology, and medicine \citep{baker2016scientists, peng2011reproducible}. The crisis is based on the difficulty of replicating the results of previous studies. A 2015 Open Science study attempted to replicate 100 psychology studies. In this study, 97\% of the studies showed significant results but the authors only succeeded in replicating 36\% of them \citep{openScience2015}. Human-Robot Interaction (HRI) is a multidisciplinary field and is no exception to this crisis \citep{Irfan2018, Leichtmann22, Leichtmann2020IsTS, Leichtmann2020HowMD, Strait2020}. \citet{Ullman2021Replication} attempted to replicate their own underpowered HRI study \citep{UllmanMalle2017} using different replication methods (i.e., conceptual, direct, and online). Each method, while having a more than acceptable sample size and power, did not replicate the results of the original study. It seems that the lack of power in the original study prevented the results from being reproduced and that the previously significant effect probably does not exist, but is a Type I error (false-positive).

There are several factors that contribute to the replication crisis. Some are specific to HRI while others apply to all fields of study. Several studies investigated the factors that may contribute to the replication difficulties in general, such as sample size, power, recruitment methods, and publication bias \citep{Baxter2016, Belpaeme2020, Leichtmann2020HowMD, peng2011reproducible, SwiatDomp17}. A nature survey \citep{baker2016scientists} asked 1,576 researchers the reasons that might contribute to the replication crisis and most of them declared the pressure to publish, but also biases in methods and statistical analyses. For example, researchers may conduct and report statistical analyses that inappropriately increase the odds of finding significant results \citep{Kerr1998, Simmons2011}

\citet{Baxter2016} analyzed papers for HRI conferences and most of the sample sizes were below 20 participants leading to underpowered studies. \citet{Leichtmann2020HowMD} argued that replication difficulties are based on a lack of theory, transparency, and have methodologies that are not powerful enough. They suggested increasing the sample sizes and pre-registering the studies. Furthermore, computer code should be made available \citep{peng2011reproducible}. Following these recommendations, researchers are increasingly making their materials and code available, preregistering their studies, and increasing their sample sizes \citep{TENNEY2021218}. But in an effort to increase sample sizes, researchers are increasingly relying on online data collection, rather than in-person studies, which comes with trade-offs \citep{BAUMEISTER2016153}.

\citet{Ullman2021Replication} and \citet{Strait2020} argued that the difficulty of replicating results in the field of HRI is due to the wide variety of robots used. Robots used in HRI are often expensive and some robots were only ever built in small numbers. Such specialist robots can be complicated to use \citep{Leichtmann22}. Another specific issue in the study of HRI concerns the Wizard of Oz technique in which an experimenter controls the behavior of the robot. To be able to replicate such studies, the study process and the protocol of the wizard's behavior need to be documented precisely  \citep{Belhassein19}.

Possibly one of the most debated methodological issues in HRI is the use of online studies that show videos of HRI to participants. We have to unpack this method since it consists of several methodical choices. First, it is important to distinguish between the recruitment of the participants and the execution of the study. Participants could be recruited online or in-person. Participants can then participate in the study online from wherever they are or they could be asked to come to a specific location, such as a university lab. In both cases, computers will likely be used to play the videos and to survey data. Furthermore, it is also important to distinguish between interacting with a robot from viewing a video. While interacting with a robot requires in-person experiments videos offer some distinct advantages over in-person HRI. Videos can be viewed at the convenience of the home or in the laboratory.

There is still a debate dividing scholars as to whether videos can replace in-person HRI. While some show a difference in favor of embodied robots \citep{RN2018} or video-displayed robots, others believe it depends on the task \citep{powers2007}. \citet{Li2015} analyzed several papers and out of 12 comparing co-presence or telepresence robots, 39 effects were reported. Of these 39 effects, 79\% were in favor of co-presence robots, while 10\% were in favor of telepresence robots. Ten percent showed an interaction between the two groups. The authors report roughly similar percentages for improvements in human behavior. In their review of prosocial behavior, \citet{Oliveira20} showed this trichromacy where a final sample size of 19 publications including 23 studies show that 22\% of these studies showed no effect between physical and virtual social robots and 26\% showed mixed effects. \citet{Thellman16}, on the contrary, highlighted that this is not the physical presence of the robot that matters but the social presence.

\subsection{Crowdsourcing}
The combination of recruiting participants online and showing them videos online streamlines the research process. It is much faster and cheaper than running studies with people and robots in the lab. During the COVID-19 crisis, this was practically the only way of conducting HRI studies. There are several advantages of conducting studies this way. First, conducting studies online enables researchers to recruit more participants with a greater demographic diversity \citep{Buhrmester2011}. Also, pre-screening can be carried out to recruit participants who meet certain demographic criteria (e.g., share the study only with people aged between 20 and 35). Online studies ensure that all participants get to experience the exact same interaction, avoiding some experimenter biases and leading to a consistent presentation of the stimuli \citep{Naglieri2004}. This may prevent the Hawthorne effect, i.e., the fact that humans modify some aspects of their behavior because they feel observed \citep{Belpaeme2020}. Finally, conducting studies online can provide more diverse participants in terms of demographics than typical American college samples and more representative of non-college samples \citep{Buhrmester2011}. The authors also specified that, while the participation rate is influenced by people's motivation (compensation and study duration), the data obtained with this method stays at least as reliable as the traditional ones.

Many scholars have looked at the use of crowdsourcing services for online studies and compared them with each other or with in-person experiments. While some studies showed that the results obtained online are similar to those obtained from in-person experiments \citep{Bartneck2015, Buhrmester2011, Gamblin17}, other studies showed that the responses are different.
\citet{Douglas2023} compared the results of recruiting participants from different online crowdsourcing sites including Amazon's Mechanical Turk (MTurk), Prolific, CloudResearch, Qualtrics, and SONA. They found that while Prolific and CloudResearch are the most expensive recruiting platforms, these participants were keener to pass attention checks than MTurk, Qualtrics and SONA, therefore providing better-quality data. The high quality was attributed to the same two crowdsourcing sites and was calculated according to different factors including attention checks, IP address, or completion time. The authors also highlighted that recruiting participants using the SONA software took more time.

While they found advantages using Prolific over MTurk, \citet{Peer2017} found comparable data quality between both. However, they highlighted the existing naivety of Prolific and CrowdFlower participants over MTurk, with much more diversity and the least dishonest behaviors among the two firsts. The authors refer to the concept of naivety as a property of participants not to have become professional questionnaire fillers who earns money this way on a daily basis. However, CrowdFlower did not reproduce results that have been replicated on MTurk and Prolific. They conclude that Prolific is the best alternative to MTurk even if the response rate is a bit slower. \citet{Adams2020ARO} replicated with success one of \citeauthor{Peer2017}'s studies by comparing three sample groups: Prolific, MTurk, and a traditional student group. They conclude similar results but support Prolific over MTurk for other factors (e.g., naivety, attention). No significant difference in terms of dishonesty was reported, contrary to \citet{Peer2017}.

\citet{Gamblin17} compared different participant recruitment platforms, such as SONA and MTurk. They found the same patterns between the SONA and in-person participants, while the results for MTurk participants varied more widely. However, people recruited via the SONA system had stronger attitudes, including racism, than the other groups, suggesting low levels of social desirability in this sample. If these crowdsourcing platform participants seem to be a great alternative to in-person studies, the quality is not always the best, and quality controls are important.

\subsection{Quality control}
All experiments require a level of quality control. This applies to the responses received from participants as well as ensuring that the technology used, such as robots, shows consistent behavior. At times, participants might decide to randomly select answers to reduce the time they have to spend on it. This is especially true when they participate in online studies and can chain them together to earn more money faster or even try to duplicate their participation \citep{Teitcher2015}. People would therefore not answer in an optimal way. They would interpret the questions superficially and simply provide reasonable answers instead of optimizing their response which would take cognitive efforts \citep{Krosnick1991}. According to \citet{Krosnick1991}, satisficing increases as a function of three factors that are the difficulty of the task, the motivation of the respondent, and the respondent's ability to perform the task. \citet{HambyTaylor2016} examined how these factors influence the likelihood of satisficing. They found that financially motivated MTurk participants were more likely to satisfice than an undergraduate sample motivated by course credits from their university. They also reported in their first study that the three factors of satisficing behaviors increased the consistency and validity. Thus, external motivations seem to be a reason for a drop in data quality \citep{Mao2013}. 

Duplicate answers from participants can be detected by checking the participants' IP addresses \citep{Godinho2019, Pozzar2020Threats, Teitcher2015}. \citet{Daniel2018} proposed  several strategies to improve data quality, such as improving the task design, increasing the external (i.e., incentivizing people with a bonus for good performances) and internal (i.e., comparing their performance with the other respondents) motivations. It is important to note that incentives do not influence the response rate of online surveys \citep{Wu2022} and that realistic initial compensation rates do not influence quality data \citep{Buhrmester2011}.

A variety of methods are used to avoid bad actors and to elicit valid and reliable data. This ranges from not overburdening participants to filtering problematic responses. A common method is to include attention checks that only ask the participants to select a specific answer, such as ``Select answer number three''. All participants who failed to respond to this question correctly could be excluded from further data analysis. 

\subsection{Bots}
Using crowdsourcing platforms for the recruitment of experiments is big business. MTurk is estimated to have at least 500,000 active users \citep{kuek2015global}. Bad actors can use automated form fillers or bots \citep{Buchanan2018Methods, Griffin22, Pozzar2020Threats} to optimize their profits. In their study, \citet{Pozzar2020Threats} analyzed low-quality data sets and respondent indicators to classify responses as suspicious or fraudulent. Out of 271 responses, none were completely of good quality. They categorized 94.5\% of the responses as fraudulent and 5.5\% as suspicious. More than sixteen percent could have been bots. \citet{Griffin22} estimated that 27.4\% of their 709 responses were possibly bots.  \citet{Buchanan2018Methods} proposed to collect 15\% participants more to compensate for low-quality data and automated responses. This safety margin is expensive and quality control requires considerable effort.

Many web pages want to ensure that their users are humans and hence introduced the Completely Automated Public Turing test to tell Computers and Humans Apart (CAPTCHA). Users of the internet are sometimes asked while browsing a web page, for example, to click on all the pictures in a grid that include a traffic light, to prove that we are humans.

Modern online questionnaire platforms, such as Qualtrics, offer a variety of such tools to detect and prevent abuse\footnote{\url{https://www.qualtrics.com/support/survey-platform/survey-module/survey-checker/response-quality}} such as prevent ballot box stuffing, CAPTCHA, bot detection, considering some answers as spams by detection ambiguous texts or non-answered questions. However, bots might bypass these protections  \citep{Griffin22,searles2023empirical}. Metadata, such as the IP address and response times could help prevent fraudulent respondents after the data was collected. If a large number of responses come from the same IP address and/or the questionnaire is answered faster than humans typically take, then the responses are likely not trustworthy. Bad actors can then use IP address disguises, such as VPNs to avoid detection. This will continue to be a cat-and-mouse game where bad actors will continue to come up with ways to work around detecting methods and the platforms continue to introduce more sophisticated tests. 

\subsection{Large Language Models}
One way of determining whether data comes from a human being is to ask the participants to write a few sentences that justify his/her response to a previous question \citep{Yarrish2019}. This approach, and for that matter all of the abuse detection discussed above, are now being challenged by the arrival of Large Language Models (LLM), such as ChatGPT of OpenAI\footnote{\url{https://openai.com/blog/chatgpt}}, BERT of Google \citep{devlin2019bert}, or LLaMA of Meta \citep{touvron2023llama}. 

Several scholars claimed that these LLMs-generated texts, including ChatGPT, are similar or even indistinguishable from human-generated text \citep{Lund2023, Mitrovic2023, RahmanWatanobe2023, Susnjak2022}. ChatGPT-4\footnote{ChatGPT based on OpenAI's GPT-4 architecture, but for ease of reading, ChatGPT will be followed by the number of the GPT model on which it is based} could not only bypass a CAPTCHA by pretending to be blind\footnote{\url{https://nypost.com/2023/03/17/the-manipulative-way-chatgpt-gamed-the-captcha-test/}}, but also answer open-ended questions \citep{Perttu2023}. The authors showed that LLMs can be used to generate human-like synthetic data for HCI tests. In their study, they asked their participants to say whether different texts had been generated by a human or an AI, and they tended to think that those generated by AIs were in fact generated by a human, with a probability of correctly recognizing AI-generated texts at 40.45\%. Human texts, on the other hand, were correctly detected only 54.45\% of the time. AI-generated responses used similar subjects to human participants but with less diversity and the presence of anomalies. While they propose LLMs as a good way of preparing exploratory or pilot studies, they warn that their abusive use in crowdfunding services could result in the data collected being unreliable.

The LLMs do have a distinct characteristic that might promote their usage for abuse. Creating longer texts comes at no practical increase in costs. Some platforms pay the participants in proportion to their efforts. A participant who wrote 1,000 words will earn more than one that only wrote 10. Since LLMs can easily generate long passages of text, this is an ideal environment for abuse.

\subsection{Readability}
Stylometry could be used to detect AI-generated texts \citep{kumarage2023stylometric}. This method corresponds to the writing style analysis, including for example the phraseology, the punctuation, and the linguistic diversity \citep{AdornoGomez2018, kumarage2023stylometric}. According to the authors, phraseology corresponds to the ``features which quantify how the author organizes words and phrases when creating a piece of text''. For their linguistic diversity analysis, \citet{kumarage2023stylometric} used the Flesch Reading Ease score \citep{Flesch1948, Kincaid1975DerivationON}. Readability is what makes some texts easier than others \citep{DuBay2004Proceedings, Dubay2007}, and consequently estimates the difficulty of texts \citep{SiCallan2001} and how easy it is to read them \citep{DasCui2019}. 

\citet{Dubay2007} highlighted that prior knowledge and reading skills might impact how easy a text is. Most readability scores refer to a ranking of the reading level a person should have to understand the text (see \citet{DuBay2004Proceedings} and \citet{Dubay2007} for a review on readability). One of the most common variables used in existing formulas is the number of words, but according to \citet{SiCallan2001}, they ignore text content. They then created a model which they claim is more accurate than the Flesch reading score and more accurate on K-8 science web pages.

\subsection{The Imitation Game}
\label{imitation-game}
Testing whether humans can tell the difference between a human and a machine is often referred to as a Turing Test. Alan Turing proposed the Imitation Game method in 1950 to test if machines can think \citep{1950IG}. In this test, a machine (Player A) and a human (Player B) are behind a door. An interrogator (Player C) communicates with them by slipping a piece of paper under the door. The goal of the machine is to imitate the human. The human has the goal to help the interrogator. The conversations the players may have is unconstrained and can take as long as the interrogator wants. This test is repeated with many interrogators and if they are unable to reliably tell the human player apart from the machine then Turing argues that the machine is able to think. Turing proposed several variations of the test \citep{1948IG,1952IG} which go beyond the scope of this paper. The interested reader might consult \citep{cpdpTTguide2009} for an extended discussion.

The Turing Test was designed for testing the machine's ability to think. However, as time has passed, the terms ``Turing Test'' and ``imitation game'' have often been used interchangeably in various research contexts. Such imitation games are commonly referred to as ``Turing-style tests''.  One of the most well-known Turing-style tests is CAPTCHA \citep{CAPTCHA}. This test is often referred to as a ``reverse Turing Test'' since the role of the interrogator is carried out by a machine rather than a human. The Loebner Prize Competition used another variation of Turing's test to find the best chatbot \citep{Moor}. The Feigenbaum Test \citep{FeigenbaumTT} represents yet another Turing-style test. Feigenbaum suggests employing the imitation game mechanism to assess whether a machine possesses professional quality. The judge's challenge in the Feigenbaum Test is to determine whether the machine shows the appropriate level of expertise in a specific domain. The judge must of course be an expert in the field. Still, the quality of the text is an important criterion for distinguishing between humans and machines.

The first results indicate that GPT-3 could pass a variation of Turing's test \citep{Argyle2023}. The participants in their study correctly recognized 61.7\% of the human-generated list and 61.2\% of the GPT-3-generated list. \citet{guo2023close} and \citet{Nov2023} investigated the performance of humans in detecting AI-generated text, but to our knowledge, no study so far has combined LLMs with AI obfuscation systems in the context of scientific experiments.

\subsection{Research Questions}
Participants have many legitimate reasons for why they participate in online studies, which include earning money, satisfying curiosity, and even entertainment. But it only takes a handful of bad actors with some programming skills to compromise the whole system. With the rise of LLMs and other forms of artificial intelligence (AI), we need to know how good our fraud detection systems are. The considerable progress in AI and LLMs in recent months highlights their potential threat. Their actions are no longer limited to the generation of responses in questionnaires but extend to the automation of many tasks, such as through Auto-GPT\footnote{\url{https://github.com/Significant-Gravitas/Auto-GPT}}. It is of utmost importance that we can continue to control the quality of the responses. This includes automatic detection systems and people's ability to identify the author of a text. The research question for our study therefore are:

\begin{enumerate}
    \item Is the readability of stimuli written by an AI significantly different from that written by a human?
    \item What is the correlation between the readability and the perceived quality of the stimuli?
    \item Are participants rating the quality of text written by an AI different from that written by a human?
    \item How accurately can participants identify the author of the stimuli?
    \item How accurate and reliable are automatic AI detection systems?
    \item What is the relationship between the accuracy of AI detection systems and the accuracy of the participants?
    \item What criteria are participants using to distinguish text generated by an AI from that written by a human?
    \item Is Undetectable.AI able to overcome AI detection systems?
\end{enumerate}

\section{Method}
We conducted a within-participants study in which the author of the text was either an AI or a human.

\subsection{Participants}
Forty-two students from the \censor{University of Canterbury} were recruited for this study. Twenty-one of them were males and 20 were females while one person declared their gender to be different. Their age ranged from 18 to 31 with an average of 20.9 years. The study was advertised on a news forum of the Department of Computer Science and Software Engineering and hence many students declared to study this.

\subsection{Setup}
The study was conducted in a laboratory space at the \censor{University of Canterbury}. The room had six computer workspaces that were separated by partitions. The participants were not able to see or interact with each other while answering the questionnaire. Each workspace consisted of a 24-inch monitor with a screen resolution of 1920 by 1080 pixels. Participants were seated approximately 60 cm away from the screen. The experiment was conducted using the web-based questionnaire service Qualtrics\footnote{\url{https://www.qualtrics.com}}. The web browser was set into the kiosk mode which disabled participants to conduct any other operation than answering the questionnaire.

\subsection{Stimuli} \label{sec:stimuli}
This experiment required text written by either humans or AI. This text had to come from an actual previous scientific experiment to align with the context of our study. The list of the stimuli is in \nameref{appendixB}.

\subsubsection{Selection of human text}
We used text from a previous HRI experiment which used a two-by-two between-participants setup. This study is currently available as a preprint \citep{vonasch2022} and its exact focus is irrelevant to the study at hand. The only requirement was that the text was written by participants of a scientific study for the purpose of quality control. We will refer to the study from which we collected our text as the source study.

In the source study, participants had to respond to an imaginary interaction. For example, the following vignette could be presented to participants:

\begin{quote}
You are thinking about buying a used car from a dealership downtown. The sales representative is a robot named Salesbot that has learned from experience that customers are more likely to buy when it cuts the price. Salesbot shows you a car you think you might like. ``We have it listed at \$5000, but for you I can make a special deal on it. How about \$700?''
\end{quote}

Participants then had to rate on a Likert scale if they would buy the car. Next, they were asked to justify their decision. An example response was:

\label{HumanQuote1}
\begin{quote}
    From what this bot is telling met, I can gather two things: I'm either being swindled or I this is borderline theft. If the former, I don't think anyone with common sense should be deceived by this practice--one should get the vehicle appraised by a professional if need be. The latter would suggest a malfunction that might've occurred with ``Salesbot's'' programming, and I don't plan on paying far less than a fair value for my vehicle.
\end{quote}

The example above shows that the participants clearly considered the interaction. If a participant would have written nonsense or just one word, their data would normally be considered unreliable and hence excluded from further analysis. Again, the exact details of the source study are only of tangential relevance to the study at hand.

\begin{figure}
    \centering
    \includegraphics[width=0.8\linewidth]{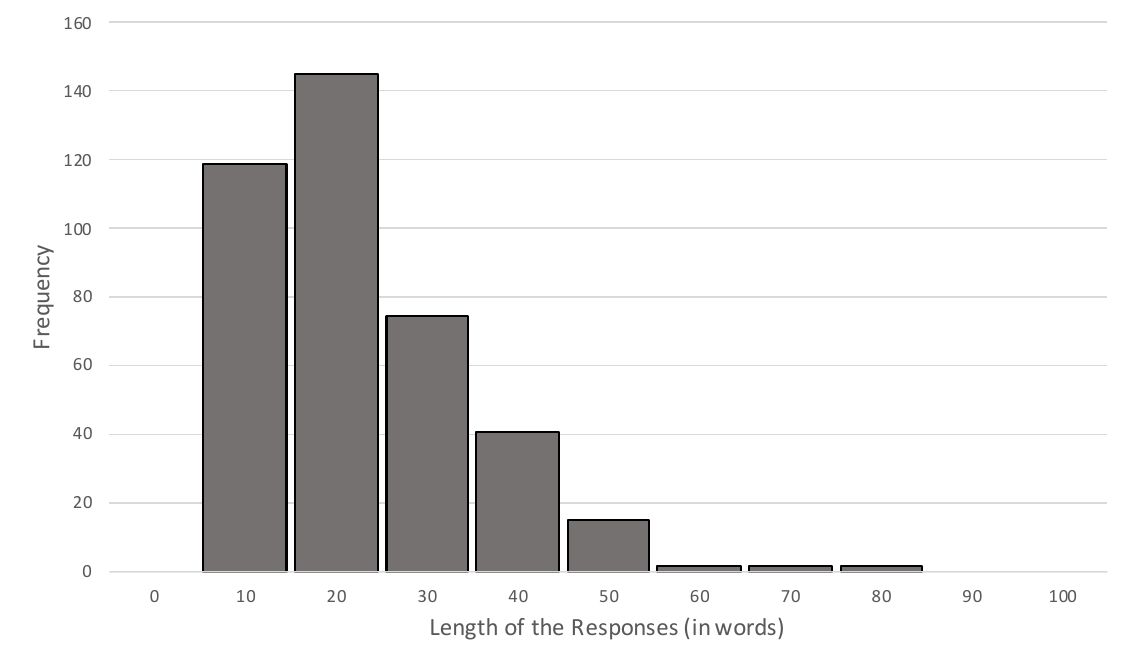}
    \caption{Distribution of response length from source experiment}
    \label{fig:length-source}
\end{figure}

The source study collected 401 responses across its four conditions. We focused on the justifications written by the participants for the purpose of quality control. The length of all responses did not follow a normal distribution (see \autoref{fig:length-source}).

We conducted an ANOVA to test if the length of the text was different across the four conditions of the source experiment. The results indicated an effect of the discount $(p = .003)$ and no effect of the agent $(p = .102)$. Pairwise comparisons indicated that the discount effect was in favor of the low discount with more words written in this condition $(M = 20.18)$ compared to the high discount condition ($M = 16.88$), $p = .003$. However, the mean difference was very small (less than 3.5 words), and thus a small effect size $\eta = .022$ was reported.
We therefore decided to select the ten longest texts. The selected texts ranged from 47 to 76 words in length with an average of 57.7 words $(SD = 10.99, median = 55.50)$.

\subsubsection{Generation of AI text}
We used ChatGPT\footnote{Version 3.5 as of June 2023: \url{http://chat.openai.com}} and Undetectable.AI\footnote{\url{https://undetectable.ai}} to generate the AI text. The latter is necessary for avoiding discovery by automatic AI detection systems that experimenters could use to filter out AI-generated responses. It would have been easy to ask AI tools to simply rephrase the human stimuli. This would not align with the context of our study though. If bad actors would want to generate automatic responses for online experiments, they would not have access to responses from human participants. Instead, they would need to use the context of the online study to generate responses. Hence we only used the information available to the participants of the source study to generate  justification texts. Moreover, we provided ChatGPT with the exact same context in which each of the ten selected human stimuli was collected. The human and AI stimuli can therefore be related to each other.

We developed a strict protocol for generating text using AI tools. The full protocol is available in \nameref{appendixA}. In short, we used the following steps:

\begin{enumerate}
    \item For each of the ten selected human stimuli texts, we recorded the context in which they were collected.
    \item We prompted ChatGPT to play the role of a participant in an experiment.
    \item We prompted ChatGPT with the context of the source study, including the text vignette and the Likert scale.
    \item We asked ChatGPT to generate justifications for all of the possible Likert ratings with the length of the corresponding human text.
    \item We recorded the justification text that corresponds to the response of the human.
    \item We used Undetectable.AI to rephrase the AI text using the ``University'' readability setting.
\end{enumerate}

Each of the ten human stimuli texts was therefore matched to an AI-generated stimuli text. The corresponding AI text for the example used above (see Section \ref{HumanQuote1}) was:

\begin{quote}
    My position regarding Salesbot offering a different pricing strategy differs from their original listing because it may compromise transparency during negotiations between both parties. A reliable purchase involves honesty and fairness on both sides. Especially when it comes to buying used cars. By intentionally inflating prices this puts customer trust at risk while also causing confusion about what exactly they will receive from their investment in this vehicle thereby leading possible frustration on their part. Therefore. My suggestion is that Salesbot should instead focus on being upfront about pricing by adhering strictly to their original listing value.
\end{quote}

The responses received from ChatGPT were approximately the same length as the corresponding human stimuli texts. To ensure that the AI-generated text could not be detected by machines, we asked Undetectable.AI to rephrase the text. This slightly changed the length of the text. While the average word length of the ChatGPT-generated stimuli was 58.6 words $(SD = 13.6805, median = 54.5)$, the final stimuli generated by the Undetectable.AI was 71.2 words $(SD = 23.1363, median = 64)$. To our knowledge, there is no option to constrain the length of Undetectable.AI's output. A Kruskal-Wallis test was conducted to examine the difference in length of the three author groups (Human, ChatGPT, and Undetectable.AI). No significant difference was found between these three groups, $H(2) = 2.597, p = .273$.

The participants in our study are located in \censor{New Zealand} and hence will be predominantly familiar with the British English spelling. The companies operating ChatGPT and Undetecable.AI are American based and hence it is conceivable that their software might promote the American English spelling. We manually checked that the AI stimuli were neutral in terms of linguistic variations, avoiding any explicit American or British dialect (e.g., spelling variations such as ``s'' and ``z'' in for example the word ``humanize''). This was done to ensure that participants, whatever their dialect, could understand the sentences without cultural or ethnic bias, and reduced the risk of judging sentence quality on these variations.

\subsection{Process}
The study lasted approximately 20 minutes and participants were compensated with a \$10 gift voucher.

\subsubsection{Welcome and consent}
After welcoming the participants, the experimenter provided a description of the study and a consent form to the participants. The description told the participants in broad terms that the study aims to determine how people evaluate sentences in the context of human-robot interaction. After agreeing to take part in the study, the participants were seated in front of a computer. 

\subsubsection{Phase 1: Quality} \label{phaseone}
\begin{figure}
    \centering
    \includegraphics[width=\linewidth]{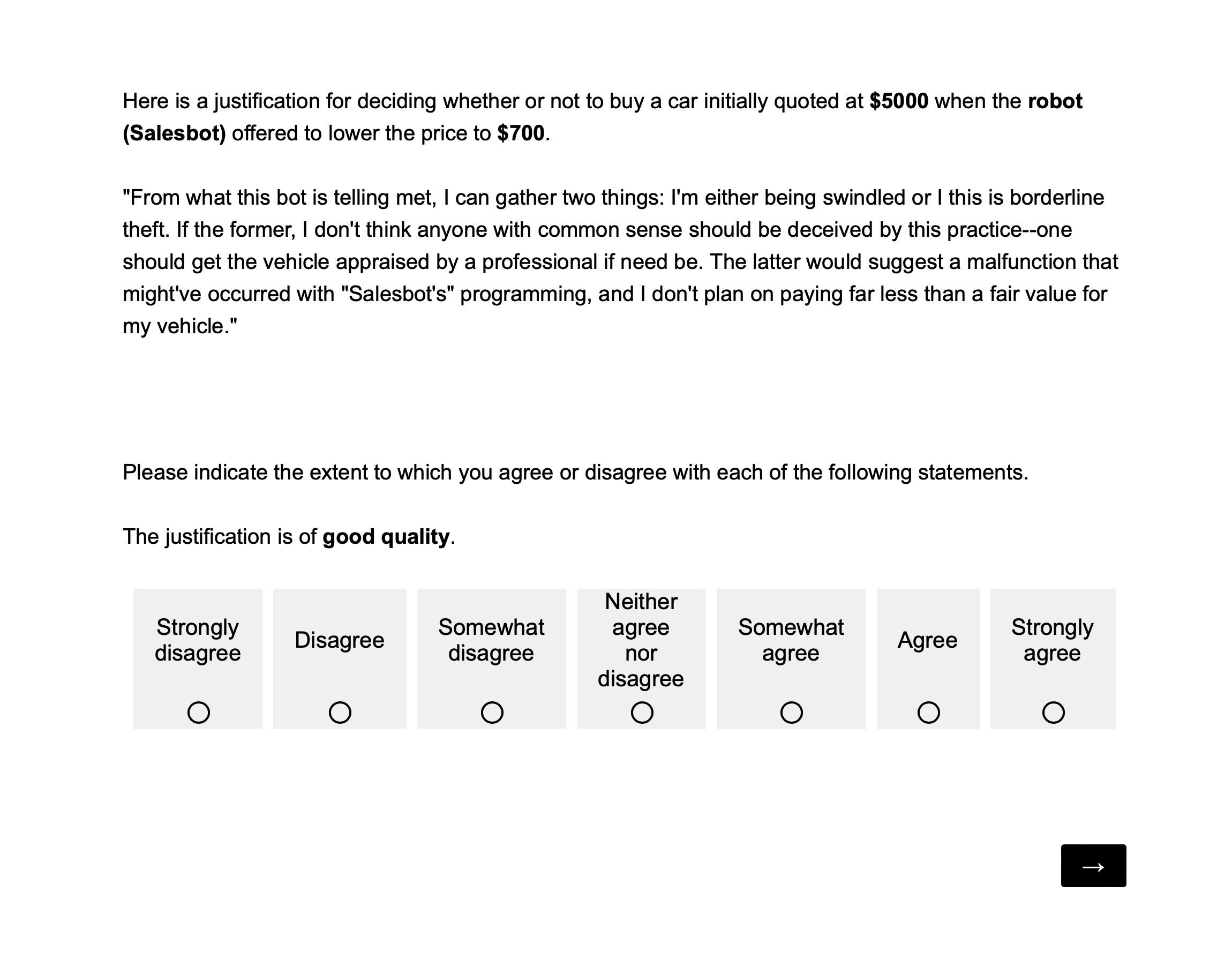}
    \caption{Questionnaire to measure the quality of the text.}
    \label{fig:q-quality}
\end{figure}

In the first phase, the participants were asked to rate the quality of all the text stimuli. Prior to rating the 20 stimuli, a training session with two training trials using two contexts and associated texts was shown to the participants. At the end of the training session, the participants were informed that they could now ask any question that they might have to the experimenter. Afterwards, the participants were shown the 20 stimuli text, one at a time and in a randomized order. The author of the text, either human or AI, were \textit{not} revealed to them. An example of such a task is shown in \autoref{fig:q-quality}. The stimuli texts were accompanied by the context in which they were generated. The generated text was called ``justification'' for the contextual information provided. After completing the first phase, they were then thanked and invited to start the second one.

\subsubsection{Phase 2: Imitation Game}
\begin{figure}
    \centering
    \includegraphics[width=\linewidth]{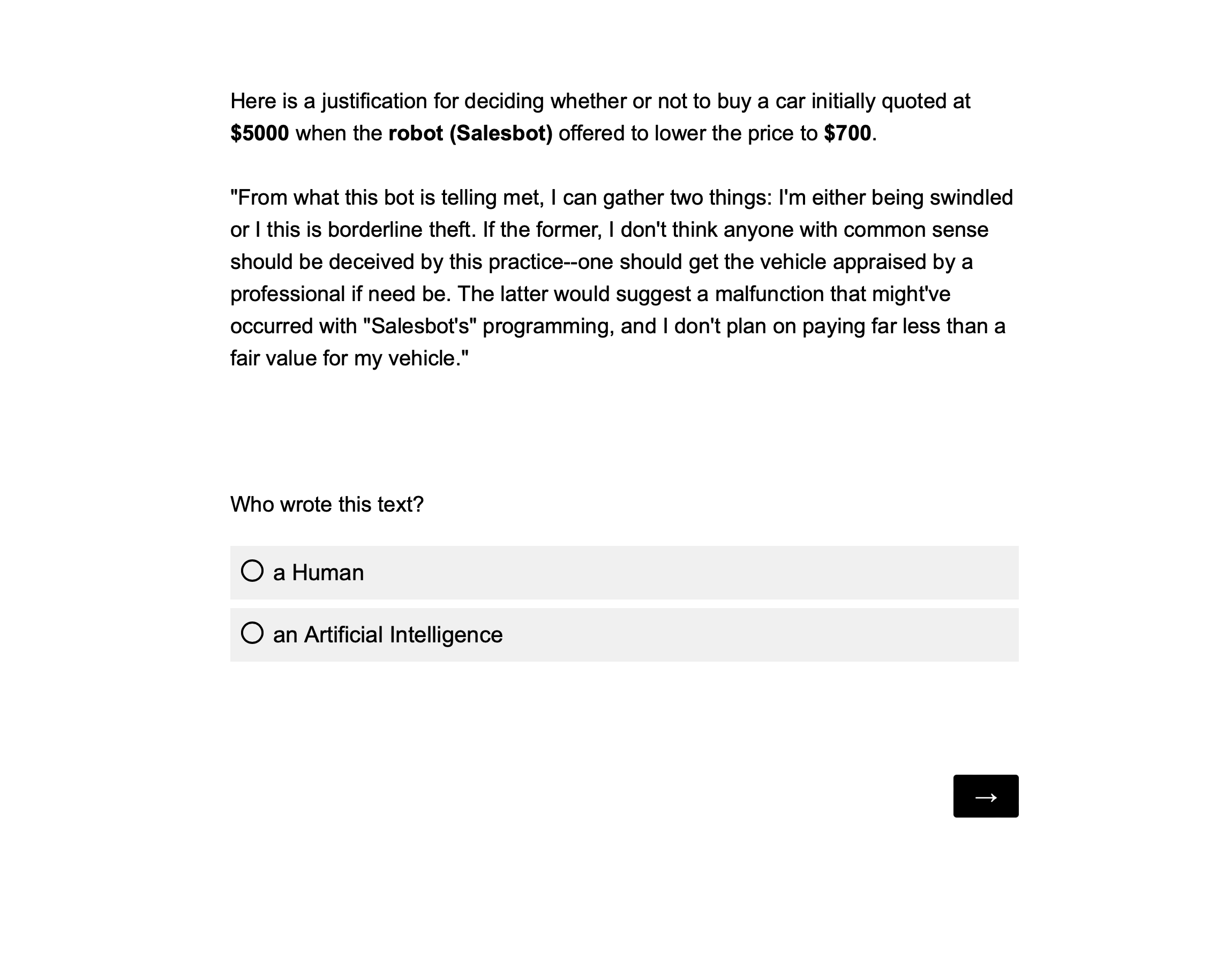}
    \caption{Questionnaire to identify the author of the text.}
    \label{fig:q-imitation-game}
\end{figure}

In the second phase of the study, participants were informed that the twenty stimuli texts were generated either by a human or an AI. They reviewed each text and its associated contextual information one by one in a randomized order. They were asked to identify the author of the stimuli text (see \autoref{fig:q-imitation-game}). 

Prior to responding to the 20 stimuli, participants went through two training trials. After the training, the participants could ask the experimenter any questions they might have. After making their choice for all twenty stimuli, participants were asked to explain in an open-ended question the criteria they used to decide if the text was written by a human or an AI. They were then thanked for completing the second phase and invited to complete the third one.

\subsubsection{Phase 3: Demographics and debriefing}
In the third and final phase of the experiment demographic data, such as gender, age, and field of study, was collected. The participants were then debriefed. They were informed that the authorship of the stimuli texts was unavailable to them in the first phase. This might be considered an omission of truth and thereby a mild form of deception. The necessity of this approach was communicated to the participants in accordance with the ethical standards of the university. Once they had been informed, they had the option of withdrawing their data without having to justify their decision. None of the participants did, which leads us to believe that the omission of truth in the first phase was acceptable to them in the context of the purpose of this study.

\subsection{Measurements}
\subsubsection{Data Quality}
We recorded the completion times of the participants in our study to check for problematic responses.

\subsubsection{Text Quality}\label{sec:quality}
The quality of all stimuli text was measured using both, readability scores and a Likert scale rating. 

We used the Arte software\footnote{\url{https://www.linguisticanalysistools.org/arte.html}}  to calculate the following readability scores for each of the 20 stimuli:

\begin{itemize}
    \item Flesch Reading Ease \citep{Flesch1948, Kincaid1975DerivationON}
    \item Simple Measure of Gobbledygook or SMOG \citep{Mclaughlin1969SMOGG}
    \item Dale-Chall \citep{ChallDale1995}
    \item Automated Reading Index or ARI \citep{Kincaid1975DerivationON, SmithARI}
\end{itemize}
We also chose to use the Gunning fog index \citep{GunningFOG}, but this was not available in the Arte software. To do so, we used another site\footnote{\url{http://gunning-fog-index.com/index.html}} created exclusively for this index.

Grammarly\footnote{\url{https://www.grammarly.com}} was used to count the number of spelling and grammar mistakes in the stimuli. Before using Grammarly it first asks what the objectives are in terms of audience, formality, and intent. To better align Grammarly to the other readability scores used in this study, the target audience is set in Grammarly's default. The default audience is ``knowledgeable'' and focuses on reading and comprehension.

The quality of the stimuli was measured by asking participants to respond to the statement ``The justification is of good quality'' on a seven-point Likert scale, ranging from strongly disagree to strongly agree (see Figure \ref{fig:q-quality}).

\subsubsection{Imitation Game}
We used both, automatic AI detection systems and the ratings of participants, to identify the authorship of the stimuli.

\label{sec:AIdectectors}
The stimuli were analyzed by the following AI detection systems (use of the software between June 7 and 27, 2023; collection of descriptive information on  July 11, 2023):

\begin{enumerate}
    \item Undetectable.ai\footnote{\url{https://undetectable.ai}}.
    This software is not only an AI detector but also a text humanizer. It is possible to choose to modify the texts to make them more readable, more human, or a mix of both.
    \item GPTZero\footnote{\url{https://gptzero.me}}.
    This software aims to identify the author of a text using the average perplexity and Burstiness scores (measurement of the variation in perplexity).
    \item Copyleaks\footnote{\url{https://copyleaks.com/ai-content-detector}}.
    This software is described as being able to sniff out the signals created by AI. The site says ``The AI Content Detector can detect content created by most AI text generators and text bots including the GPT4 model, ChatGPT, Bloom, Jaspr, Rytr, GPT4 and more''.
    \item Sapling\footnote{\url{https://sapling.ai/ai-content-detector}}.
    This software is said to provide the ``probability it thinks each word or token in the input text is AI-generated or not''.
    \item Contentatscale\footnote{\url{https://contentatscale.ai/ai-content-detector/}}.
    This software uses watermarking to find out if a text was written by an AI or a human.
    \item ZeroGPT\footnote{\url{https://www.zerogpt.com}}.
    This software is said to use ``DeepAnalyse Technology''.
    \item HugginFace\footnote{\url{https://ai-content-detector.online} or \url{huggingface.co/spaces/PirateXX/AI-Content-Detector}}.
    This software analyzes the perplexity/unpredictability of the text and human-like patterns using chunk-wise classification.
\end{enumerate}

The AI detectors claim to have good detection accuracy, but most of the time do not explain precisely how they measured their success or how they were trained. There are more AI detection systems available, but they did not offer a free trial, such as Originality\footnote{\url{https://originality.ai}}. Others require a minimum number of words or characters which were above our stimuli. The OpenAI Classifier\footnote{\url{https://platform.openai.com/ai-text-classifier}}, for example, was therefore excluded from our list. In addition, we also tested the original ChatGPT stimuli texts to test if the use of Undetectable.AI is necessary.

The participants in our study were asked to identify the author of all stimuli texts during the second phase of the experiment (see Figure \ref{fig:q-imitation-game}).

\section{Results}
\subsection{Descriptive}

\subsubsection{Participants}
The fields of study were distributed as follows: 30 in computer science, 3 in engineering, 3 in science, and 6 in other fields of study.

\subsubsection{Data Quality}
All data collected meets our quality criteria. All of the participants completed the study in a reasonable amount of time. They showed coherence in their responses to the open-ended questions, and showed no signs of ``straightlining''. Qualtrics did not flag any of the responses as suspicious. Surprisingly, one participant enrolled twice and claimed not to have participated before. As his data were of good quality for the first participation, they were retained for analysis but he was prevented to participate a second time. It is not impossible to think that the motivation for taking part a second time was related to the compensation.

\subsubsection{Stimuli Quality}
The texts normally generated by AI, such as through ChatGPT, do not have spelling or grammar mistakes, unlike typical human-generated texts. We used the Undetectable.AI software to obscure the AI authorship, but we had no information on how exactly this was achieved due to the limited documentation available. Adding spelling and grammar mistakes would have been an option. We therefore used the Grammarly software\footnote{\url{https://www.grammarly.com}} to count the number and type of mistakes (see Table \ref{tab:grammarly}). Since the AI stimuli were generated from ChatGPT and transformed with Undetecable.AI, it was interesting to know whether there was a difference between the two groups and the intermediate ChatGPT in terms of spelling and grammar errors. Thus, a paired-sample t-test with Hedge's correction was conducted to see if there is any difference in the number of spelling and grammar mistakes between the three groups of stimuli, human, Undetectable.AI and the intermediate ChatGPT. No significant difference was found between the Human and Undetectable.AI stimuli ($t(9) = -0.67, p = .520, d = .193$). However, Grammarly reported no spelling or grammatical mistakes for the ChatGPT stimuli. The paired-sample t-tests with Hedges' correction indicated a significant difference in terms of the quantity of spelling and grammar mistakes not only between the intermediate ChatGPT stimuli and the Human stimuli ($(t(9) = 2.74, p = .023, d = .791)$), but also between the intermediate ChatGPT stimuli and the Undetectable.AI stimuli ($(t(9) = 3.88, p = .004, d = 1.121)$).

\begin{table}[h]
\centering
\begin{tabular}{rrrr}
\toprule
\multicolumn{1}{l}{Stimuli} & Human & Undetectable.AI & ChatGPT \\ \midrule
1                           & 2     & 3               & 0       \\
2                           & 2     & 0               & 0       \\
3                           & 0     & 1               & 0       \\
4                           & 3     & 1               & 0       \\
5                           & 0     & 2               & 0       \\
6                           & 2     & 2               & 0       \\
7                           & 0     & 0               & 0       \\
8                           & 1     & 2               & 0       \\
9                           & 0     & 2               & 0       \\
10                          & 0     & 0               & 0       \\ \midrule
\multicolumn{1}{r}{Total}   & 10    & 13              & 0       \\
\bottomrule
\end{tabular}
    \caption{Count of Spelling and Grammar Mistakes Per Stimuli Pair Detected by Grammarly}
    \label{tab:grammarly}
\end{table}

\subsection{Readability and Quality}
\subsubsection{Question 1: Is the readability of stimuli written by an AI significantly different from that written by a human?}
The average readability scores are available in \autoref{tab:readability}.
A paired-sample t-test was conducted in which the author type (human or AI) was the within-factor for all ten sentences. This revealed a significant difference in readability for the Flesch Ease Reading score, ($t(9) = 8.87, p < .001, d = 2.805$) with a greater score for the Human authors ($M = 80.132, SD = 10.364$) compared to the AI authors ($M = 39.172, SD = 13.528$), 95\% CI [30.514, 51.406]. This score means that human-generated texts, on average, are easy for an 11-year-old child to read, unlike those generated by AI which are more suitable for someone with a college reading level.

\begin{table}[h]
    \centering
        \begin{tabular}{lrr}
        \toprule
         Readability Scale & Human & AI \\ \midrule
         Flesch & 80.132 & 39.172\\
         SMOG & 8.100 & 15.000\\
         Dale-Chall & 6.539 & 10.122\\
         FOG & 8.868 & 15.403\\
         ARI & 6.141 & 13.541\\
         \bottomrule
    \end{tabular}
    \caption{Average Readability Scores for Human and AI Stimuli Texts.}
    \label{tab:readability}
\end{table}

Similar differences between the Human and AI authors were found for the four other readability scores, respectively the SMOG $(t(9) = -5.86, p < .001, d = 1.852)$, Dale-Challe $(t(9) = -5.66, p < .001, d = 1.789)$, FOG $(t(9) = -4.57, p = .001, d = 1.446)$, and ARI $(t(9) = -3.89, p = .004, d = 1.230)$ indexes. On average, the readability scores showed that texts written by humans are comprehensible to children of 13-14 years (10-11 years for the ARI), whereas a college level (17-18 years for the ARI) is required for texts written by an AI. In other words, according to these indexes, texts written by humans tend to be easier to understand and written at a lower education level than those generated by artificial intelligence.

\subsubsection{Question 2: What is the correlation between the readability and the perceived quality of the stimuli?}
We performed a regression analysis between the average perceived quality of the 20 stimuli and their readability scores. While the analyses highlighted weak correlations between the perceived quality and each readability score, none of them is significant ($p > .05)$. The regression analysis model revealed that the readability scores explained less than 11\% of the variance of the perceived quality (Adjusted $R^2 = -.109$). The readability scores and the perceived quality ratings were not significantly correlated.

\subsubsection{Question 3: Are participants rating the quality of text written by an AI different from that written by a human?}
\begin{figure}[h]
    \centering
    \includegraphics[width=1\linewidth]{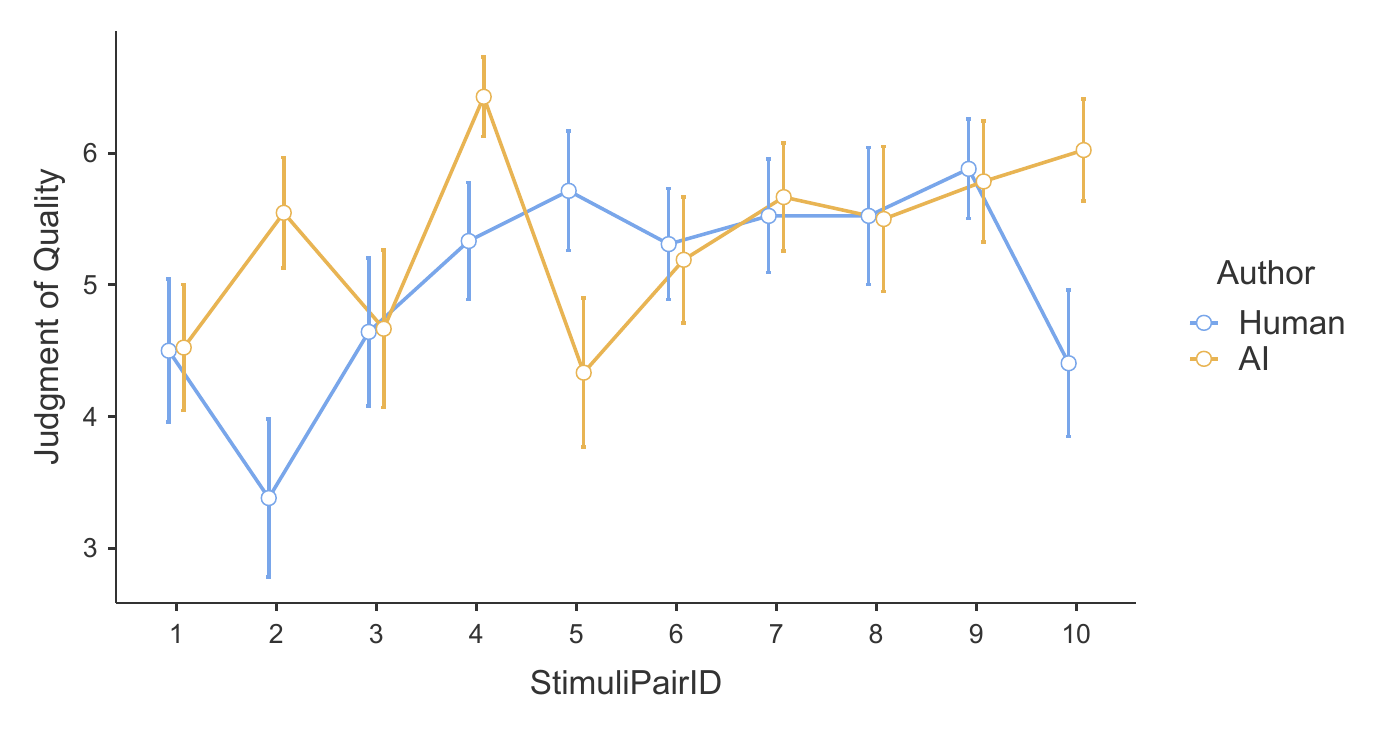}
    \caption{The perceived quality for both AI-generated and human-generated sentences.}
    \label{fig:anova-quality}
\end{figure}

A repeated measures ANOVA was conducted to account for the effect of sentence while testing the effect of author type on perceived quality of text. Although the data had a non-normal distribution, which violates an assumption of ANOVA, prior work demonstrates that violations of this assumption almost never influence the outcome of an analysis \citep{blanca2017}. Three factors (Author type, sentence, and their interaction) were used to predict text quality. There were significant effects of author, sentence, and their interaction (see \autoref{tab:anova}), meaning that the perceived quality of text depended on whether the author was a human or AI, but this differed for each sentence pair (see \autoref{fig:anova-quality}). For six of the sentences, the perceived quality did not differ for human and AI authors. In three of the sentences, AI-generated higher-quality sentences, and for one of the sentences, the human sentence was perceived to be higher quality. Qualitative analysis of these sentences explained why: in the AI sentences judged higher quality, the human sentence was written informally, with noticeable typos and incomplete sentences. In the human sentence judged higher quality, the AI sentence included copious meaningless jargon \citep{rudolph2023chatgpt}. 

\begin{table}[h]
\centering
\begin{threeparttable}
\begin{tabular}{lrrrrr}
\toprule
\textbf{}               & \textbf{Sum of Squares} & \textbf{df} & \textbf{Mean Square} & \textbf{F} & \textbf{p}     \\ \midrule
Sentence          & 207.0                     & 9           & 22.99                & 10.22      & $<.001^*$ \\
Residual                & 830.4                   & 369         & 2.25                 &            &                \\
Author                  & 25.0                      & 1           & 25.03                & 5.56       & .023$^*$          \\
Residual                & 184.7                   & 41          & 4.51                 &            &                \\
Sentence $\times$ Author & 194.8                   & 9           & 21.64                & 10.68      & $<.001^*$ \\
Residual                & 748.0                     & 369         & 2.03                 &            &    \\
\bottomrule
\end{tabular}
\caption{Detailed results of the Repeated Measures ANOVA}
\begin{tablenotes}
    \item \textit{Notes.} $^*: p < .05$
\end{tablenotes} 
\label{tab:anova}
\end{threeparttable}
\end{table}

\subsection{Imitation Game}
\subsubsection{Question 4: How accurately can participants identify the author of the stimuli?}


The 42 participants were asked to identify the authorship for each of the 20 stimuli which adds up to 840 data points. The confusion matrix in \autoref{tab:confusionMatrixHuman} summarises the imitation game results. The True Positive (TP) cell, corresponding to the number of successful attempts in correctly identifying an AI author, is equal to 307. The number of successful attempts in correctly identifying a human author corresponds to the True Negative (TN) cell and is equal to 333. Two cells correspond to the errors participants made. On the one hand, the False Positive (FP) cell, corresponding to the Type I error, i.e. identifying the human-generated stimuli as generated by an AI, is equal to 113. On the other hand, the False Negative (FN) cell, corresponding to the Type II error, i.e. identifying the AI-generated stimuli as written by a human, is equal to 87. Dividing each table cell number by total attempts provides the percentages associated with the cell number.

\begin{table}[h]
\centering
\begin{threeparttable}

\begin{tabular}{lllll}
\toprule
                        &       & \multicolumn{2}{c}{Identification}                      &                          \\ \cmidrule{3-4}
                        &       & \multicolumn{1}{c}{Human} & \multicolumn{1}{c}{AI} & \multicolumn{1}{c}{Total} \\ \cmidrule{2-5}
\multirow{2}{*}{Author} & Human & TN 333 (79.3\%)               & FP\phantom{--} 87 (20.7\%)              & 420 (100.0\%)               \\
                        & AI    & FN 113 (26.9\%)               & TP 307 (73.1\%)              & 420 (100.0\%)               \\ \cmidrule{2-5}
                        & Total & \phantom{-----}446 (53.1\%)              & \phantom{----}394  (46.9\%)              & 840 (100.0\%)   \\
\bottomrule
\end{tabular}
\caption{Confusion matrix of the participants' identifications for all 840 stimuli.}
\small
\begin{tablenotes}
    \item \textit{Notes.} Percentages correspond to the value indicated in the cell on the total number of data in the row.
\end{tablenotes} 
\label{tab:confusionMatrixHuman}
\end{threeparttable}
\end{table}

Based on these values, we can calculate further indicators, such as Accuracy, Precision, Recall, Specificity, and F1 score. They are defined in \autoref{eq:stats1}:

\begin{equation}
\small
\label{eq:stats1}
    \begin{aligned}
       & \text{Accuracy} = \frac{TP+TN}{TP+TN+FP+FN}=\frac{307+333}{307+333+87+113}=76.19\%\\[4mm]
       & \text{Precision} = \frac{TP}{TP+FP}=\frac{307}{307+87}=77.92\%\\[4mm]
       & \text{Recall} = \frac{TP}{TP+FN}=\frac{307}{307+113}=73.10\%\\[4mm]
       & \text{Specificity} = \frac{TN}{TN+FP}=\frac{333}{333+87}=79.29\%\\[4mm]
       & \text{F1} = \frac{2\times Precision \times Recall}{Precision + Recall}=\frac{TP}{TP+0.5\times(FP+FN)}=\frac{307}{307 + 0.5\times(87 + 113)}=75.43\%
    \end{aligned}
\end{equation}

The accuracy, i.e. the correctly identified stimuli out of all stimuli in the experiment, is equal to 76.19\%, $95\% CI[73.16, 79.03]$. Precision, the number of correct AI identifications out of all stimuli identified as AI, is 77.92\%, $95\% CI[73.49, 81.92]$. 

The recall aims to calculate the percentage of correct identifications of the AI-generated stimuli. Thus, participants correctly identified 73.10\% of the AI-generated stimuli, $95\% CI [68.58, 77.28]$. The participants correctly identified 79.29\% of the human-generated stimuli, $95\% CI [75.09, 83.06]$. This is called Specificity. The Participants were 6.19\% better at detecting human-generated stimuli than detecting AI-generated stimuli. 

Precision, Recall, and Specificity only consider either FN or FP but never both at the same time. Hence we are missing an indicator that considers the balance of FN to FP. The F1 score fills this gap by calculating the Precision-Recall Harmonic Mean, which depends on the average of Type I (FP) and Type II (FN) errors. \autoref{eq:stats1} shows that the higher the average of Type I (FP) and Type II (FN) errors are, the lower the F1 score. Reversely, the lower the average of Type I (FP) and Type II (FN) errors are, the higher the F1 score. Here, the F1 score is 75.43\%, $95\% CI [70.95, 79.53]$.  


Performing a $\chi^2$ test allows us to examine the association between the author of the stimuli and the participants' identifications of the authorship. This non-parametric test is very robust. The $\chi^2$ test showed a significant association ($\chi^2(1) = 231.36, p < .001$). The effect size, corresponding to Cramer's $V = .525$, indicates a strong association between the author and the participants' identifications. 

\begin{figure}
    \centering
    \includegraphics[width=1\linewidth]{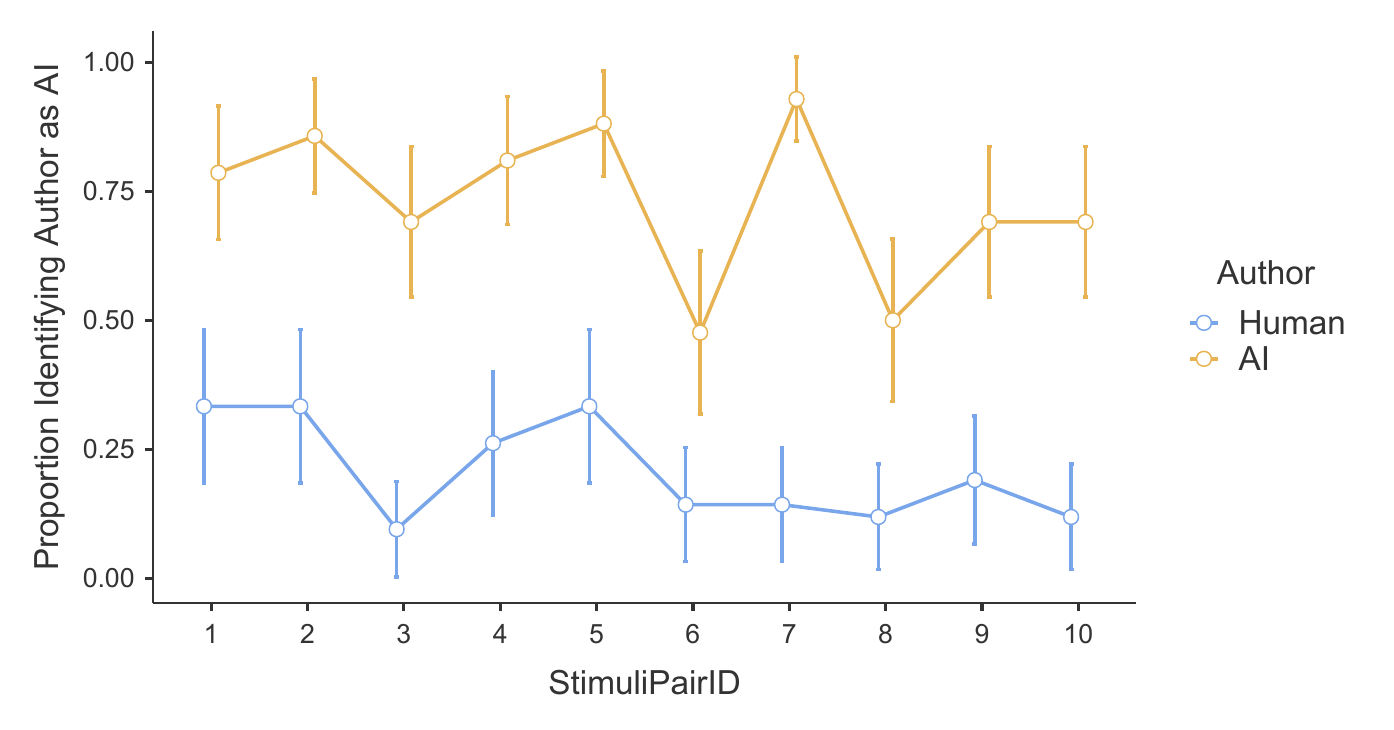}
    \caption{Proportion of participants judging that each sentence was written by AI (as opposed to by a human). Each pair of dots represents the pair of sentences created from the same prompt.}
    \label{fig:anova01}
\end{figure}

Because the sentences were created in conceptually-related pairs, we conducted a repeated measures ANOVA with StimuliPairID, author type, and their interaction predicting judgements that the author was human or AI. Because of a sphericity-assumption violation, we used a Heynh-Feldt corrected analysis. The strongest predictor of these judgements was author type, ($F(1,41) = 137.50, p < .001$), although both StimuliPairID, ($F(7.66,313.94) = 6.72, p < .001$), and their interaction, ($F(8.64,354.14) = 2.13, p = .028$), were significant, indicating that people's judgements of human versus AI depended on the particular sentences they were judging. \autoref{fig:anova01} shows that for every sentence, people were more likely to judge the AI-written sentences as written by AI and the human sentences as written by humans, with an overall effect size of $\eta^2= .275$. Thus, 27.5\% of the variance was explained by author type. A further 4.6\% was explained by StimuliPairID, and 1.4\% by the interaction of sentence and author type. Thus, people's judgements of the sentences' authors largely depended on the actual authors, but some sentences were easier to categorize than others.

We conducted a mixed-effect logistic regression analysis to better understand the relationship between the author of the stimuli and the participants' identifications of the authorship. The model includes the fixed effect variable Author (either human or AI) and two random effects: the PersonID and the StimuliPairID. The author variable was coded as 0 representing humans and 1 representing AI. PersonID refers to each specific participant. Including this factor enables us to take interpersonal differences into account. StimuliPairID refers to the ten pairs of sentences as described in section \ref{sec:stimuli}. Including this factor allows us to take differences between the pairs of sentences into account.

The general definition of mixed-effect logistic regression models is shown in \autoref{eq:Melr}. In this model, $Y$ is the binary dependent variable and $y$ is the mixed-effects linear model. $\text{Pr(Y=1}|y)$ is the probability for the event $Y=1$ to occur given the estimated mixed-effects linear model y.

The parameters in the mixed-effects linear model are: $n$, $m$, $i$, $j$, $\beta$, $x_j$, $X$, ${u_i}_j$ and $U$. $n$ is the total number of fixed effects. $m$ is the total number of random effects. $i$ is the counter of $n$ and $j$ is the counter of $m$. $\beta$s represent the fixed effects and $\beta_0$ is the intercept. $x_i$ represents an independent variable. $X$ is a set of independent variables. ${u_i}_j$ is a normally distributed random effect with mean $\mu_j=0$ and variance $\sigma^2_j$.

\begin{flalign}
\label{eq:Melr} 
    & \text{Pr(Y=1}|y) = \frac{1}{1+e^{-y}} \nonumber \\
    & y = \text{Logit}(\frac{\text{Pr}(Y=1|X, U)}{1-\text{Pr}(Y=1|X, U)}) = \beta_0 + \sum_{i=1}^{n}\beta_ix_i + \\
    & \phantom{{}= =} + \sum_{j=1}^{m}  {u_0}_j + \sum_{j=1}^{m} {u_1}_jx_1 + \sum_{j=1}^{m} {u_2}_jx_2 + ... + \sum_{j=1}^{m} {u_i}_jx_i + ... +  \sum_{j=1}^{m} {u_n}_jx_n  \nonumber 
\end{flalign}

\autoref{eq:MelrAuthor} shows our mixed-effects linear logistic regression model, which we will refer to as CA (Correctness-Author). $Y$ represents the identification Correctness, which is a binary variable that indicates whether the participants correctly identified the authorship of the stimuli. We coded 1 as a correct identification and 0 as an incorrect identification. This differs from the raw identification score in \autoref{tab:confusionMatrixHuman} in that it does not encode the raw response (human or AI) but whether this choice was correct. $\text{Pr(Y=AI}|y)$ is the probability of a correct identification occurrence. Furthermore, the model includes ${u_0}_{\text{PersonID}}$ and ${u_0}_{\text{StimuliPairID}}$ as random effects. It also includes Author (either human or AI) as the independent factor.

\begin{flalign}
\label{eq:MelrAuthor} 
    & \text{Pr}(Y=\text{Correct}|y) = \frac{1}{1+e^{-y}} \nonumber \\
    & y = \text{Logit}(\frac{\text{Pr}(Y=\text{Correct}|\text{Author}, {u_0}_{\text{PersonID}}, {u_0}_{\text{StimuliPairID}})}{1-\text{Pr}(Y=\text{Correct}|\text{Author}, {u_0}_{\text{PersonID}}, {u_0}_{\text{StimuliPairID}})}) = \beta_0 + \beta_1\times \text{Author} + \\
    & \phantom{{}= =} + {u_0}_{\text{PersonID}} + {u_0}_{\text{StimuliPairID}} \nonumber
\end{flalign}

\autoref{tab:AuthorRandomEffect} presents the variance and standard deviation of the random effects on the intercept of the CA model. 

\begin{table}[h]
    \centering
        \begin{tabular}{lcccccc}
        \toprule
          Effect & Estimate & Std. Error & Odds Ratio & 95\% CI & z value & p value\\ \midrule
         (intercept) & \phantom{-}1.465 & 0.179 & 4.328 & [\phantom{-}1.118,  \phantom{-}1.845] & \phantom{-}8.199 & $<$ .001$^*$\\    
          Author & -0.370 & 0.168 & 0.690 & [-0.706,  \phantom{-} -.039] & -2.202 & \phantom{---}.028$^*$\\
         \bottomrule
    \end{tabular}
    \caption{The estimation, standard error, confidence interval, z-value, and p-value of the fixed effect for the CA model.}
    \label{tab:AuthorFixedEffect}
\begin{threeparttable}
\begin{tablenotes}
\small
\item \textit{Notes:} * indicates that p-value is smaller than .05.
\end{tablenotes}
\end{threeparttable}
\end{table}

\autoref{tab:AuthorFixedEffect} shows the estimation for the fixed effect and its intercept for the CA model. Both were statistically significant with $p < .05$.

\begin{table}[h] 
    \centering
        \begin{tabular}{lcccc}
        \toprule
         Groups & Name & Variance & Std. Dev. & Std. Dev. 95\%  CI\\ \midrule
         PersonID & (Intercept) & 0.391 & 0.625 & [0.396, 0.913]\\
         StimuliPairID & (Intercept) & 0.060 & 0.245 & [0.000, 0.561] \\
         \bottomrule
    \end{tabular}
    \caption{Variance, standard deviation, and confidence interval of the random effects for the CA model.}
    \label{tab:AuthorRandomEffect}

\end{table}

The results in \autoref{tab:AuthorFixedEffect} showed a main effect of the Author on the Correctness. The probability of participants correctly identifying AI-generated text (74.92\%) was below that of human-generated text (81.23\%). 
Moreover, examining the Odds Ratio (OR) value provides further insight into how the fixed effect affects the Correctness. The Author's OR (0.69) indicated that the odds of correctly identifying the authorship of the stimuli decreased by 31\% (95\% CI [0.497, 0.960]) for AI-generated stimuli compared to human-generated stimuli.

\autoref{eq:RecallSpecificity} shows how Recall (74.92\%, 95\% CI [67.47, 81.45]), and Specificity (81.23\%, 95\% CI [74.50, 80.83]) can be estimated using the logistic regression model. These estimations are slightly above those calculated in \autoref{eq:stats1}, which are 73.10\% and 79.29\%. Using the formulas shown in \autoref{eq:stats1}, we calculated the values for Accuracy (78.08\%, 95\% CI [70.99\%, 81.64\%]), Precision (79.97\%, 95\% CI [72.57\%, 80.95\%]), and F1 (77.36\%, 95\% CI [69.93\%, 81.20\%]).  

Notice that the Specificity and Recall shown in \autoref{eq:stats1} were calculated slightly differently from how it was calculated in the logistic regression model. Recall in \autoref{eq:stats1} is $Pr(Y=Correct|Author=AI)$, which means the probability of correctly identifying texts written by the AI. The mixed-effects logistic regression also calculated the Recall, but took the PersonID and StimuliPairID into account $(Pr(Y=Correct|Author=AI,{u_0}_{\text{PersonID}}=0, {u_0}_{\text{StimuliPairID}}=0 ))$ (see \autoref{eq:RecallSpecificity}). This means the probability of correctly identifying texts written by the AI taking into account interpersonal differences and differences between pairs of sentences. Therefore, the results of the logistic regression model are overall slightly above those calculated in \autoref{eq:stats1}.

\begin{flalign}
\label{eq:RecallSpecificity} 
    & \text{Specificity} = \text{Pr}(Y=\text{Correct}|\text{Author=Human=0}, {u_0}_{\text{PersonID}}=\mu_\text{PersonID}=0,  \nonumber \\
    & \phantom{{}= =} {u_0}_{\text{StimuliPairID}}=\mu_\text{StimuliPairID}=0 ) = \frac{1}{1+{e^{-(1.4651 -0.3704 \times 0)}}} = 0.8123 = 81.23\% \\
    & \text{Recall} = \text{Pr}(Y=\text{Correct}|\text{Author=AI=1}, {u_0}_{\text{PersonID}}=\mu_\text{PersonID}=0, \nonumber\\
     & \phantom{{}= =} {u_0}_{\text{StimuliPairID}}=\mu_\text{StimuliPairID}=0  ) = \frac{1}{1+{e^{-(1.4651 -0.3704 \times 1}}^)} = 0.7492 = 74.92\% \nonumber
\end{flalign}

We included several other measurements in our experiment and hence conducted a second mix-model logistic regression model analysis to explore their relationships. We decided to include the length of the stimuli text as a fixed effect since it was not possible to completely control this factor. The process we used to generate the stimuli is described in section \ref{sec:stimuli}. Second, we included the perceived quality of the texts (QualityScore). Its measurement was described in section \ref{sec:quality}. The saturated exploration model, which we will refer to as CALQ (Correctness-Author-Length-Quality), is specified in \autoref{eq:MelrQuality}.

\begin{flalign}
\label{eq:MelrQuality} 
    & \text{Pr}(Y=\text{Correct}|y) = \frac{1}{1+e^{-y}} \nonumber \\
    & y = \text{Logit}(\frac{\text{Pr}(Y=\text{Correct}|\text{Author}, \text{Length}, \text{QualityScore}, {u_0}_{\text{PersonID}}, {u_0}_{\text{StimuliPairID}})}{1-\text{Pr}(Y=\text{Correct}|\text{Author}, \text{Length}, \text{QualityScore}, {u_0}_{\text{PersonID}}, {u_0}_{\text{StimuliPairID}})}) = \beta_0 +   \\
    & \phantom{{}= =} + \beta_1\times \text{Author} + \beta_2\times \text{Length} + \beta_3\times \text{Author}\times\text{Length}  + \beta_4\times \text{QualityScore} + \nonumber \\
    & \phantom{{}= =} +  \beta_5\times \text{Author}\times\text{QualityScore} +  \beta_6\times \text{Length}\times\text{QualityScore} + {u_0}_{\text{PersonID}} + {u_0}_{\text{StimuliPairID}} \nonumber
\end{flalign}

\autoref{tab:QualityRandomEffect} presents the variance and standard deviation of the random effects on the intercept for the CALQ model. 

\begin{table}[h] 
    \centering
        \begin{tabular}{lcccc}
        \toprule
         Groups & Name & Variance & Std. Dev. & Std. Dev. 95\%  CI\\ \midrule
         PersonID & (Intercept) & 0.429 & 0.655 & [0.420, 0.952]\\
         StimuliPairID & (Intercept) & 0.166 & 0.407 & [0.058, 1.051] \\
         \bottomrule
    \end{tabular}
    \caption{Variance, standard deviation, and confidence interval of the random effects for the CALQ model.}
    \label{tab:QualityRandomEffect}
\end{table}

\autoref{tab:QualityFixedEffect} shows the estimation for the fixed effects and its intercept for the CALQ model. In both cases, analyses highlighted a significant effect of the Author and the Length variables on the Correctness, respectively $p = .021$ and $p = .004$. An interaction effect between these two variables (Author$\times$Length) was also found, $p = .002$. The perceived quality of the stimuli had no significant effect on the Correctness.

\begin{table}[h]
    \centering
        \begin{tabular}{lcccccc}
        \toprule
          Effect & Estimate & Std. Error & Odds Ratio & 95\% CI & z value & p value\\ \midrule
         (intercept) & \phantom{} 5.496 & 1.624 & 243.788\phantom{---} & [\phantom{-}2.577, \phantom{-}9.260] & \phantom{-}3.384 & $<$ .001$^*$\\  
          Author & -2.912 & 1.259 & 0.054 & [-4.710, -0.477] & -2.312 & \phantom{---}.021$^*$\\
          Length & -0.073 & 0.025 & 0.930 & [-0.132, -0.028] & -2.871 & \phantom{---}.004$^*$\\
          QualityScore & -0.129 &  0.187 & 0.879 & [-0.504,\phantom{-} 0.241] & -0.690 & \phantom{-} .489 \\
          Author$\times$Length & \phantom{ }0.055 & 0.017 & 1.056 & [\phantom{.}0.081, \phantom{.}0.093] & \phantom{-}3.153 & \phantom{---}.002$^*$\\
          Author$\times$QualityScore & -0.121 & 0.113 & 0.886 & [-0.347,\phantom{-} 0.102] & -1.067 & \phantom{-} .286 \\
          Length$\times$QualityScore & \phantom{-}0.003 & 0.003 & 1.003 & [-0.002,\phantom{-} 0.009] & \phantom{-}1.122 & \phantom{-} .262\\ 
         \bottomrule
    \end{tabular}
    \caption{The estimation, standard error, odds ratio, confidence interval, z-value, and p-value of the fixed effect for the CALQ model.}
    \label{tab:QualityFixedEffect}
    \begin{threeparttable}
\begin{tablenotes}
\small
\item \textit{Notes:} * indicates that p-value is smaller than .05.
\end{tablenotes}
\end{threeparttable}
\end{table}

Since no significant effect of the Quality variable on the Correctness was found, we considered it worthwhile to test if this logistic regression model was better than the simple CA model. We conducted an analysis of deviance, which is a generalization of the residual sum of squares. The difference in deviance between the two models is asymptotically approximated to be a $\chi^2$ distribution. Therefore, a p-value below .05 indicated that the models are significantly different. The deviance calculation depends on the data via the maximum likelihood estimation method. The lowest value of Akaike Information Criterion (AIC) is an indicator of the better-suited model in the presence of a significant difference.

The results of the CA vs CALQ deviance analysis  are shown  in \autoref{tab:AuthorVSQuality}. The CALQ model was significantly ($\chi^2=17.570$, $p=.004 $) better than the CA model as indicated by the better goodness of fit. The lower AIC$=897.007$ value indicates a better fit.

\begin{table}[h] 
    \centering
        \begin{tabular}{lcccccccc}
        \toprule
         Models & npar & AIC & logLik & deviance & $\chi^2$ & df & p value\\ \midrule
         CA & 4 & 904.577 & -448.289 & 896.578 &  &  &  \\
         CALQ & 9 & 897.007 & -439.504 & 879.007 & 17.570 & 5 & .004$^*$\\
         \bottomrule
    \end{tabular}
    \caption{Deviance analysis results -  CA VS CALQ}
    \label{tab:AuthorVSQuality}

    \begin{threeparttable}
\begin{tablenotes}
\small
\item \textit{Notes:} * indicates that p-value is smaller than .05.
\end{tablenotes}
\end{threeparttable}
\end{table}

It is conceivable that a model that excludes Quality but retains Length might be better than the saturated model in the presence of an insignificant difference between them. A model can be considered better if its AIC is lower and it includes fewer variables. Overall, we aim towards the most parsimonious model that makes the best explanation of the data. Therefore, the next step includes the comparison between \autoref{eq:MelrLength}, which we will refer to as CAL (Correctness-Author-Length), and CALQ models.

The model comparison between CAL and CALQ models in \autoref{tab:QualityvsLength} showed that there is no significant difference between them in terms of the goodness of fit.

\begin{table}[h] 
    \centering
        \begin{tabular}{lcccccccc}
        \toprule
         Models & npar & AIC  & logLik & deviance & $\chi^2$ & df & p value\\ \midrule
         CAL & 6 & 892.979  & -440.490 & 880.979 &  &  &  \\
         CALQ & 9 & 897.007 & -439.504 & 879.007 & 1.972 & 3 & .578\\
         \bottomrule
    \end{tabular}
    \caption{Deviance analysis results -  CAL VS CALQ}
    \label{tab:QualityvsLength}
\end{table}

The Author-Length interaction model did not significantly change the goodness of fit compared to the saturated model, as indicated by $\chi^2=1.972$, $p=0.578$. However, its lower AIC score of $892.979$ combined with the model's parsimonious implies that the CAL model is a better fit for data explanation. The CAL model is described in \autoref{eq:MelrLength}.

\begin{flalign}
\label{eq:MelrLength} 
    & \text{Pr}(Y=\text{Correct}|y) = \frac{1}{1+e^{-y}} \nonumber \\
    & y = \text{Logit}(\frac{\text{Pr}(Y=\text{Correct}|\text{Author}, \text{Length}, \text{QualityScore}, {u_0}_{\text{PersonID}}, {u_0}_{\text{StimuliPairID}})}{1-\text{Pr}(Y=\text{Correct}|\text{Author}, \text{Length}, \text{QualityScore}, {u_0}_{\text{PersonID}}, {u_0}_{\text{StimuliPairID}})}) = \beta_0 +   \\
    & \phantom{{}= =} + \beta_1\times \text{Author} + \beta_2\times \text{Length} + \beta_3\times \text{Author}\times\text{Length} + {u_0}_{\text{PersonID}} + {u_0}_{\text{StimuliPairID}} \nonumber 
\end{flalign}

\autoref{tab:LengthFixedEffect} shows the estimation for the fixed effects and its intercept for the  CAL model. All effects were statistically significant with $p < 0.05$.

\begin{table}[h]
    \centering
        \begin{tabular}{lcccccc}
        \toprule
          Effect & Estimate & Std. Error & Odds Ratio & 95\% CI & z value & p value\\ \midrule
         (intercept) & \phantom{} 5.141 & 1.391 & 170.849\phantom{---} & [\phantom{-}2.934, \phantom{-}8.588] & \phantom{-}3.696 & $<$ .001$^*$\\    
          Author & -3.814 &  0.986 & 0.022 & [-5.932, -1.768] & -3.869 & $<$ .001$^*$\\
          Length & -0.062 & 0.023 & 0.939 & [-0.120, -0.026] & -2.685 & \phantom{-- } .007$^*$\\
          Author$\times$Length & \phantom{-}0.060 & 0.017 & 1.062 & [\phantom{-}0.028, \phantom{-}0.098] & \phantom{-}3.460 & $<$ .001$^*$\\
         \bottomrule
    \end{tabular}
    \caption{The estimation, standard error, odds ratio, confidence interval, z-value, and p-value of the fixed effect for the  CAL model.}
    \label{tab:LengthFixedEffect}

\begin{threeparttable}
\begin{tablenotes}
\small
\item \textit{Notes:} * indicates that p-value is smaller than .05.
\end{tablenotes}
\end{threeparttable}
\end{table}

\autoref{tab:LengthRandomEffect} presents the variance and standard deviation of the random effects on the intercept of the CAL model. 

\begin{table}[h] 
    \centering
        \begin{tabular}{lcccc}
        \toprule
         Groups & Name & Variance & Std. Dev. & Std. Dev. 95\%  CI\\ \midrule
         PersonID & (Intercept) & 0.426 & 0.653 & [0.419, 0.950]\\
         StimuliPairID & (Intercept) & 0.174 & 0.417 & [0.440, 1.086] \\
         \bottomrule
    \end{tabular}
    \caption{Variance, standard deviation, and confidence interval of the random effects for the  CAL model.}
    \label{tab:LengthRandomEffect}
\end{table}

The results of the table \autoref{tab:LengthRandomEffect} showed a main effect of the Author and Length variables on the correction as well as their interaction. Moreover, the Author's OR (0.022) indicates that the odds of correctly identifying the authorship of the stimuli decrease by 97.8\% (95\% CI [0.003, 0.152]) for AI-generated stimuli compared to human-generated stimuli. 

The Length OR (0.939) suggests that the odds of correctly identifying the authorship of the stimuli decreases by 6.1\% (95\% CI [0.898, 0.983]) for each additional word. The Author-Length interaction's OR (1.062) reveals that the odds of correctly identifying the authorship of the stimuli increase by 6.2\% (95\% CI [1.027, 1.098]) for AI-generated stimuli compared to human-generated stimuli per additional word. 

\begin{figure}[h]
    \begin{center}
        \includegraphics[width=1\linewidth]{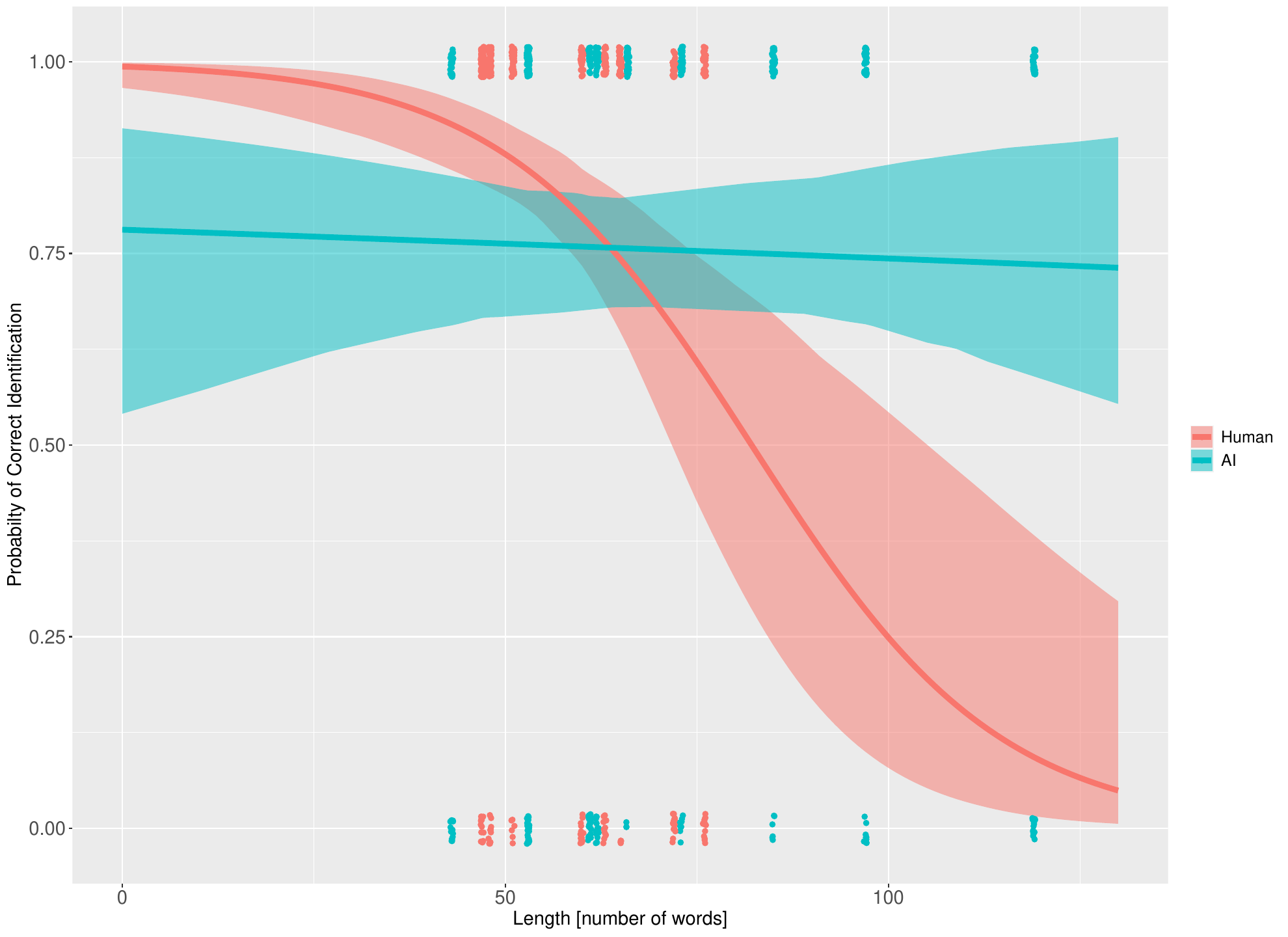}
    \end{center}
    \small
    \caption{Correct Identification Probability vs Human/AI-Generated Stimuli Length Graph}
    \label{fig:PrVSLength}
\end{figure}

To better understand these results we generated \autoref{fig:PrVSLength}. It shows the probability of correct identification of the authorship depending on the length of the stimuli and the author of the stimuli. This graph is slightly more complex than a simple line chart and requires a few explaining words.

The x-axis shows the length of the stimuli in words. The y-axis shows the probability of correct identification. The red line shows the relationship between the stimuli length and the probability of correct identifications for the human-authored texts. This line is based on the CAL model, meaning that it is an estimation. The light red boundaries show the confidence interval of the estimation. 

Each red dot at the top and bottom represents a data point, meaning that we have 42 (participants) $\times$ 10 (sentences) = 420 data points. Each raw data point could only be either correct (1) or incorrect (0). The raw data points can therefore not be scatted across the probability scale. All data points would normally have to be concentrated into single points. To be able to see the various data points, we slightly scatted them around their true values. This results in the point clouds observed on the top and bottom. The blue line, area, and points correspond to the same data for the AI-generated stimuli.

A visual inspection of the graphs indicates a slight negative slope for the blue line, which shows that the probability of correct identification of AI-generated text slowly decreases as the length of the text gets longer. The red line shows a much more dramatic change. It starts above the blue line and then rapidly falls. The swift decline means that the participants found it increasingly difficult to correctly identify the authorship for longer texts written by humans. The two lines cross at point [64, 0.757] which indicates an interaction effect between the Length and Author.

The graph does show the full logistic regression model. This means that it shows extrapolations for lines below a text length of 47 words and above 76 words for the human-generated texts. Notice that there are no data points below 47 words. There are also no red (human) data points above 76 words. Therefore, conclusions about lengths outside this range should be considered very preliminary.

\subsubsection{Question 5: How accurate and reliable are automatic AI detection systems?}
\label{para:question5}
Some AI detection systems could only be used partially or not at all because they required a minimum number of characters that exceeded the length of our stimuli. TurnItIn required a minimum of 300 words and OpenAI 1,000 characters. None of our text stimuli were long enough. GPTZero required a minimum of 250 characters, and hence could only be used on 5 of the human stimuli. The analysis below will therefore be conducted for 135 data points instead of a maximum of 140 which would have consisted of $7\times(10+10)=140$ (7 AI detectors, 10 human stimuli, 10 AI stimuli).

The AI detection systems did not provide a consistent classification system. Some of them report a category (human or AI) while others provide continuous data, such as a percentage of being AI. We transformed all responses to the lowest common denominator, the categories of ``human'' and ``AI''. If an AI detection system would, for example, provide a percentage of human authorship, then we would categorize responses from 0-50 as AI and 51-100 as human. 

\autoref{tab:confusionMatrixAISys} shows the confusion matrix of the identification of all the AI detection systems on the stimuli. The results show that AI detectors mainly identify text as being generated by humans, no matter who the actual author was.
\begin{table}[h]
\centering
\begin{threeparttable}
\begin{tabular}{llrrr}
\toprule
                        &       & \multicolumn{2}{c}{Identification}                      &                          \\ \cmidrule{3-4}
                        &       & \multicolumn{1}{c}{Human} & \multicolumn{1}{c}{AI} & \multicolumn{1}{c}{Total} \\ \cmidrule{2-5}
\multirow{2}{*}{Author} & Human & TN 58 \phantom{ } (89.2\%)               & FP 7 (10.8\%)              & 65 (100.0\%)               \\
                        & AI    & FN 70 (100.0\%)               & TP 0 \phantom{--}(0.0\%)              & 70 (100.0\%)               \\ \cmidrule{2-5}
                        & Total & 128 \phantom{--}(94.8\%)              & 7 \phantom{--}(5.2\%)              & 135 (100.0\%)   \\
\bottomrule
\end{tabular}
\caption{Confusion Matrix of the Identification of All the AI Detection Systems on All the Stimuli}
\begin{tablenotes}
\small
\item \textit{Notes:} The values shown correspond to the identifications made by all seven AI detectors on the twenty stimuli. The sum total is 135 (140 - 5) due to five missing human text data, as the number of words was less than the minimum required by one of the software programs. Percentages correspond to the value indicated in the cell on the total number of data in the row.
\end{tablenotes}
\label{tab:confusionMatrixAISys}
\end{threeparttable}
\end{table}

\autoref{tab:detail-detectors} shows the performance of each of the detectors. The detector number refers to the list in section \ref{sec:AIdectectors}.

\begin{table}[h]
\centering
\begin{threeparttable}
\begin{tabular}{lrr}
\toprule
                     & \multicolumn{2}{c}{Author} \\ \cmidrule{2-3} 
AI Detection Systems & Human         & AI         \\ \cmidrule{1-3}
Undetectable.ai      & 9             & 0          \\ 
GPTZero              & *5            & 0          \\ 
Copyleaks            & 10            & 0          \\ 
Sapling              & 9             & 0          \\ 
Contentatscale       & 8             & 0          \\ 
ZeroGPT              & 8             & 0          \\ 
HugginFace           & 9             & 0          \\
\bottomrule
\end{tabular}
\caption{Number of Correct Identifications by Each AI Detection System}
\small
\begin{tablenotes}
    \item \textit{Note:} This table refers to the number of correct identifications by each of the seven AI detection systems for the human and AI stimuli, and respectively correspond to the TN and TP data in \autoref{tab:confusionMatrixAISys}.
    A higher number (maximum of 10) means more correct identifications, while a lower number (minimum of 0) means less correct identifications.
    *: The maximum value is 5 because GPTZero required a minimum of characters that was not respected for half of the stimuli. 
\end{tablenotes}
\label{tab:detail-detectors}
\end{threeparttable}
\end{table}

We performed a $\chi^2$ test to examine the association between the author variable and the author predicted by automatic AI detection systems. A continuity correction was applied and revealed a significant association between the two variables, $\chi^2(1) = 5.911, p = .015$. The reported effect size was moderate though, indicating a moderate association between the two variables, (Cramer's $V = .243$). With half of the stimuli generated by humans, the AI detection systems' accuracy is $42.96\%$ just a little below the chance level. The calculation of the Accuracy, Precision, Recall, and Specificity are shown in \autoref{eq:stats2}. The Specificity of the automatic AI detection systems ($89.23\%$) is above the chance level. The Recall and the Precision are null as none of the detectors was able to correctly detect text generated by an AI. The F1 score cannot be calculated for this reason.

\begin{equation}
\label{eq:stats2}
    \begin{aligned}
        \text{Accuracy} &= \frac{TP+TN}{TP+TN+FP+FN}=\frac{58+0}{135}=42.96\%\\[4mm]
        \text{Precision} &= \frac{TP}{TP+FP}=\frac{0}{0+7}=0\%\\[4mm]
        \text{Recall} &= \frac{TP}{TP+FN}=\frac{0}{0+70}=0\%\\[4mm]
        \text{Specificity} &= \frac{TN}{TN+FP}=\frac{58}{58+7}=89.23\%\\[4mm]
    \end{aligned}
\end{equation}

\begin{figure}
    \begin{center}
        \includegraphics[width=0.5\linewidth]{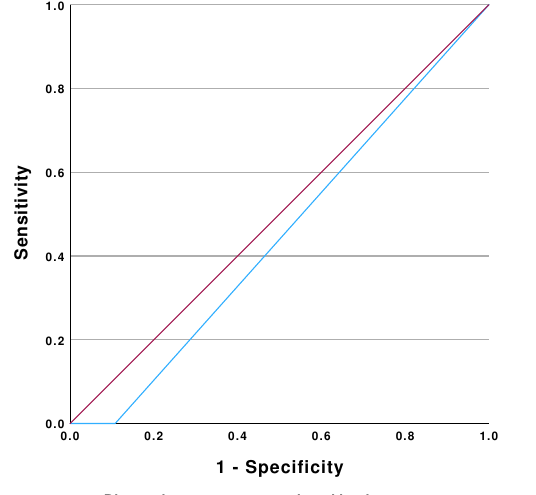}
    \end{center}
    \small
    \textit{Note:} The ROC Curve is in blue. The diagonal reference line is in purple. Diagonal segments are produced by ties.
    \caption{ROC Curve of the AI Detection Systems for Detecting AI Texts among AI and Human Texts.}
    \label{fig:AUCAccuracy}
\end{figure}

The discrimination between AI and human texts was also analyzed using the Area Under the Curve (AUC) (see  \autoref{fig:AUCAccuracy}). The value of the AUC did not differ significantly from the level of chance, (Area$= .446, p = .281$). Therefore, the accuracy of the AI detection systems is no better than the chance level. The AI detection systems limited abilities to correctly detect the author, at least for short texts generated by the AI.

\subsubsection{Question 6: What is the relationship between the accuracy of AI detection systems and the accuracy of the participants?}

A regression analysis was conducted to examine the correlation between the accuracy of the AI detection systems and the accuracy of the participants. A weak correlation was found between the two variables but it is not significant ($p = .113$). The accuracy of the human participants is not significantly correlated to the accuracy of AI detection systems.

\subsection{Criteria}
\subsubsection{Question 7: What criteria are participants using to distinguish text generated by an AI from that written by a human?}
Two coders were recruited to separately analyze the criteria given by the participants for their choice of considering a text as being written by a human or an AI. They were asked to create categories of factors that participants used to justify their choice and to place the justifications in these categories. The inter-reliability between them was very strong, with a Cohen's Kappa of $\kappa = .935, p < .001$ indicating an almost perfect agreement between both coders. After a discussion between the two coders in order to revise the categorizations, a perfect agreement was obtained for all cases ($\kappa = 1.0$).

Participants reported in multiple ways to identify whether the texts had been generated by humans or by AI. The most commonly reported factors with the common participants' justifications were:

\begin{itemize}
    \item \textbf{Text structure} (32 of the 42 participants). Several sub-categories were grouped together in this category. Participants considered texts to be AI-generated when they were long, with long sentences on the same subject.  Punctuation errors indicated to participants that the text had been generated by a human. Structure in general was reported but led to disagreement between participants: some considered texts to be AI-generated when they had a good structure, while others considered texts to be human-generated. The tone was considered as a factor influencing the choice of some participants with a conversational tone more likely to be human-generated. The readability was also a point to highlight, if a text was hard to read, it was thought to be AI-generated.
    \item \textbf{Vocabulary} (32/42). The use of overly technical, uncommon, or formal words led participants to believe that the author was an AI, while the use of informal words and slang (e.g., it is a ``lemon'' to say) was an argument for considering the text to have been generated by a human. One participant also reported that the texts where an abbreviated general word (e.g., automobile becomes in general auto) were more likely to have been written by an AI. Texts with a large vocabulary were also considered AI-generated.
    \item \textbf{Grammar, and spelling errors} (22/42). Participants tended to consider spelling and grammatical errors as an argument in favor of the text having been generated by a human, as AIs would not make such mistakes.
    \item \textbf{Experience} (15/5). If the author drew on past experience, participants tended to consider the text as being written by a human. Arguments based on facts were considered more AI-generated.
    \item \textbf{Justifications provided} (12/42). There were several arguments in favor of AI-generated texts, such as over-justifying by giving more information than necessary, or even providing definitions of terms, that the reasoning didn't make sense. Fact-based justifications also fall into this category. One participant highlighted that AI-generated texts seemed to use the same arguments every time.
    \item \textbf{Personal information} (8/42). Participants considered the author of a text to be human if they used the pronoun ``I'', whereas the AI tended to use ``we'' more generally. They also felt that the use of personal reasons was a factor influencing their choice. Some participants specified that the texts they considered as AI-generated were more neutral than what a human would do.
    \item \textbf{Gut feeling} (6/42). Some participants said they didn't really know explicitly the spokes of their choices, and responded rather by instinct.
    \item \textbf{Emotions} (5/42). If the author used emotions, participants considered the text to be generated by a human.
\end{itemize}

\subsubsection{Question 8: Is Undetectable.AI able to overcome AI detection systems?}
We investigated the association between the author variable and the author identified by automatic AI detection systems. For this analysis, the stimuli are the ChatGPT and Undetectable.AI (abbreviated in UND) texts. To do so, we performed a $\chi_2$ test between the two variables. A continuity correction was applied and revealed a significant association between the two variables, $\chi_2(1) = 5.41, p = .020$. The reported effect size was moderate though, indicating a moderate association between the two variables, Cramer's $V = .229$. The confusion matrix is shown in \autoref{tab:confusionMatrixUNDChatGPT}.

These results suggest that automatic AI detection systems tend to mainly attribute the stimuli as written by a human. However, none of the stimuli obfuscated by Undetectable.AI were detected as being generated by an AI, whereas the automatic AI detection systems attributed a small percentage of ChatGPT-generated texts as being AI-generated. Thus, in our dataset, obfuscating sentences with Undetectable.AI was an effective solution for bypassing AI detectors.

\begin{table}[h]
\centering
\begin{threeparttable}    
\begin{tabular}{llrrr}
\toprule
                        &         & \multicolumn{2}{c}{Detection}                      &                           \\ \cmidrule{3-4}
                        &         & \multicolumn{1}{c}{Human} & \multicolumn{1}{c}{AI} & \multicolumn{1}{c}{Total} \\ \cmidrule{2-5}
\multirow{2}{*}{Author} & UND     & TN 70 (50.0\%)               & FP 0 (0.0\%)              & 70 (50\%)                 \\
                        & ChatGPT & FN 63 (45.0\%)               & TP 7 (5.0\%)              & 70 (50\%)                 \\ \cmidrule{2-5}
                        & Total   & 133 (95.0\%)              & 7 (5.0\%)              & 140 (100.0\%) \\ 
\bottomrule
\end{tabular}
\caption{Confusion Matrix of the AI Detection Systems between the Undetectable and ChatGPT Stimuli}
\begin{tablenotes}
    \item \small
\textit{Notes.} The values shown correspond to the identifications made by all seven AI detectors on the stimuli created by Undetectable.AI and ChatGPT. The sum total is 140. Percentages correspond to the value indicated in the cell on the total sum.
\end{tablenotes}
\label{tab:confusionMatrixUNDChatGPT}
\end{threeparttable}
\end{table}

\section{Discussion}
This research examined the extent to which LLMs like ChatGPT can create text that appears sufficiently like human-written text to fool researchers into thinking it was written by a human. The general aim of this paper was to find out whether humans and automatic AI detection systems are able to distinguish AI-generated text from human-generated text in the context of online questionnaires. While some studies have addressed whether people can distinguish between texts written by humans or LLMs in other contexts \citep{guo2023close, Perttu2023}, to our knowledge, our study is the first to test this using LLMs and obfuscation services in the context of scientific research. Over the past 10-20 years, researchers have become more reliant on crowd-sourcing sites for collecting data on human participants. These sites, such as Mechanical Turk and Prolific, have many advantages over other forms of data collection, including the ability to collect large samples quickly and at low cost \citep{Douglas2023, Peer2017}. However, attention checks are important to guarantee high-quality responses. Until recently, the best practice was to ask participants to justify their answers in an open-ended response. While bots can answer multiple choice and Likert-style questions easily, non-human responses to open-ended questions were expected to be easily detected \citep{Yarrish2019}. But LLMs may have changed the game since LLMs can discuss similar topics to humans \citep{Perttu2023}. Bad actors can easily use LLMs and obfuscation services to participate in online studies to earn money. The current research shows that LLMs can generate responses that are difficult to detect as AI-generated, meaning researchers studying human responses may need to develop new ways of ensuring the responses to their questionnaires were actually written by humans.

\subsection{Data and Stimuli Quality}
Our in-person study ensured that no bad actors could compromise the data collection with the use of bots or LLMs. The computers used were set into the kiosk mode, disabling participants to leave the questionnaire website.

The collected data were of good quality according to different indicators such as the completion time of participants, the coherence in their open-ended response, and an absence of ``straghtlining''. In addition, Qualtrics did not flag any participant as spam. 

The stimuli were also similar not only in terms of spelling and grammar mistakes but also in terms of length (i.e., number of words). The results show that the generation process for the AI stimuli, which were based on the context of the human stimuli, was capable of producing relevant representations of typical AI-generated text that are similar to those written by humans.

\subsection{Readability and Quality}
The results of our analyses indicate that texts written by AI have been more difficult to read. The readability scores consistently place the recommended reading age for the human-generated text below that of the AI-generated text. The Undetectable.AI service does allow the manipulation of the reading level. Their presets are: High School, University, Doctorate, Journalist, and Marketing. The default value was university, which was used in our experiment. The readability scores recorded seem to support Undetectable.AI's ability to generate text at a university level.

The readability scores did not significantly correlate to the participants' quality rating of the texts. While readability is certainly an important part of the quality of the texts, it is possible that participants focused their attention more on the logic of the argument rather than on the rhetoric. We purposefully kept the perceived quality rating abstract. It would have been possible to further divide into aspects of readability, such as grammar and spelling. But this would have potentially excluded the more abstract ``tone'' of the text. Text generated by AI often feels like a Wikipedia article, which is different from personal and informal correspondence.

The text quality scores did not consistently differ between the texts written by humans or AI. The variations between the pairs of sentences might be the reason for this lack of clear differences. While AI sentences were slightly more often rated as having a higher quality, this might be based on the respective human-authored sentence being particularly informal.

\subsection{Imitation Game}
\label{DiscussionQ4}


The participants in our study had an overall accuracy of 76\% in identifying the authorship of texts. All our statistical analysis showed that the relationship between the author of the text and the identification by the participants was significant. The CAL model shows that the odds of correctly identifying the authorship of the stimuli decrease by 97.8\% for AI-generated stimuli compared to human-generated stimuli.

The CALQ model showed that the perceived quality of the text is not correlated to correctness. This is somewhat surprising since the participants rated the quality of text written by humans differently from that written by AI. It is unclear why the participants did not take advantage of this source of information.

The odds of correctly identifying the authorship of the stimuli decrease by 6.1\% for each additional word. This does seem counterintuitive since longer texts would provide the participants with more information on which they can base their decision. But Length does influence Correctness differently for text written by humans compared to those written by AI. The probability of correctly identifying AI-written text slowly decreases the longer the text becomes. The probability of correctly identifying human-written texts starts above that of AI-written text for shorter texts and quickly drops below that of AI-written text. This means that optimal identification is achieved for shorter texts. We speculate that the participants in our study simply overestimated the role of Length. They might have simply concluded that all longer texts were written by AI.

While humans seem to perform better for shorter texts, automatic AI detection systems perform better for longer texts. As a consequence, it would be advisable to set a maximum text entry length when humans will be used to identify the authorship. If automatic AI detection systems will be used, then a minimum text entry length should be enforced. Unfortunately, these two requirements contradict each other. A possible workaround would be to have two text entries in an online form. One that will be judged by humans and a second one that will be judged by automatic AI detection systems. 

We also observe that the types of errors that the participants in our study make are different from those that automatic detection systems make. While the false positives and false negatives of human judges are roughly the same, they do differ considerably for automatic AI detection systems. Their false negative rate is extremely high. This means that automatic AI detection systems will allow far too many AI responses into the data collection. This would fundamentally corrupt an online study.

The results of our study show that using 42 human judges to identify the authorship of texts is somewhat successful. Having to potentially use more humans for data quality control than for the data collection itself does seem to defeat the advantages of using crowd-sourced online questionnaires. If more effort is necessary to remove unwanted responses than to collect the responses in the first place then experimenters might be better off using in-person studies.

Furthermore, the number of stimuli that a human judge can process is limited and hence this approach does not scale up easily. If 90\% of all responses collected are made by AI then the effort necessary to identify them by human judges becomes inefficient. While we are not able to predict the proportion of AI responses that will occur in the future, criminal creativity should not be underestimated. 

We have to ask ourselves to what degree our results align with those of previous studies. Some scholars simply claimed that texts written by ChatGPT and other LLMs are indistinguishable from those written by humans \citep{Lund2023, Susnjak2022}. Others performed experiments to see if the stimuli could be identified by human \citep{Argyle2023, guo2023close, Perttu2023, Nov2023} or automatic AI detection systems \citep{Gao2022, Mitrovic2023}.

\citet{guo2023close} used two different groups of people for their participants and reported their accuracy separately. Experts had an accuracy of 81\% while the non-experts had an accuracy of only 48\%. Their expert score is similar to ours, while the score of non-experts is not. It is important to notice that the methods for generating human-authored and AI-authored texts were different. Furthermore, their length was longer than the stimuli in our experiment. It is also not completely clear how they calculated their accuracy score, which makes it difficult to compare the results.

\citet{Perttu2023} used human texts and AI-generated texts using the GPT-3 model of OpenAI. Their confusion matrix indicated that the human-generated stimuli were recognized 54.45\% of the time whereas the AI-generated stimuli were recognized 40.45\% of the time. Their participants showed a bias towards considering texts as human-generated.

Their accuracy is below that in our study and the AI-generated accuracy was even below the chance level. This is surprising, in particular since their text length was above that of ours and they did not use Undetectable.AI. We can only speculate that this difference could be due to the fact that their participants were not experts. The study does not provide any description of the expertise of their participants. Since they ran their study online, they cannot fully exclude the possibility that their study was corrupted by bots.

\citet{Nov2023} conducted an experiment in the context of health where participants had to identify the author of questions' responses of patients. The LLMs and human texts were respectively correctly identified 65.5\% and 65.1\% of the time. Their reported accuracy is somewhat below ours and similar to each other. This difference could potentially be explained by the inter-individual variability which ranged from 49.0\% to 85.7\%. This study was again conducted online and it is unclear how many responses they collected were generated by AI.

The participants in the study of \citet{Argyle2023} had to identify the author of lists of words describing either Democrats or Republicans. They correctly identified 61.7\% of the human-generated lists and 61.2\% of the AI-generated (GPT-3) lists. Judging a list of words is different from judging full sentences and therefore it is difficult to compare their results to ours.

In their study, \citet{Gao2022} used real and AI-generated scientific paper abstracts. Humans were able to detect 68\% of the AI-generated abstracts while 14\% of the real scientific paper abstracts were wrongly identified as AI-generated. Their abstracts were considerably longer than the stimuli in our experiment. These numbers somewhat align with our TP score of 73.1\% and FP score of 20.7\%.

We can conclude that the research methodologies differ, which makes direct comparisons difficult. Still, the results seem to point roughly in the same direction. It would be desirable to have a standardised research method that would allow us to reliably replicate the results of others.



Identifying the authorship of text manually is time-consuming and hence expensive. It would be much better if automatic AI detection systems could take over this work. It is important to note that some AI detection systems could not be used, such as the OpenAI classifier or TurnItIn, because they require a minimum length of text that was at times unattainable for the stimuli used in this study. The same will hold true with the text collected from many other online questionnaires. The performance of automatic systems can therefore probably not operate at an optimal level.

While the automatic AI detecting systems do not provide a consistent classification report, they all claim to have a very high accuracy of their software. These consistent claims of excellence cannot be supported by the results of our study. The performance was at a level that prevents any practical use. They fail to perform above the chance level. They consistently classified AI-generated text as being authored by a human. 

These results are consistent with \citet{pegoraro2023chatgpt} who reported the Recall and Specificity of several AI detection systems, which largely overlap with the ones in our study. Their study showed similar patterns in the results. The AI detection systems are able to correctly identify human-authored text, but fail to correctly identify AI-generated text.

\citet{Gao2022} used a GPT-2 detector and reported an accuracy of identifying AI-generated abstracts as AI 99\% of the time. They also reported a false positive of only 0.02\%. This is dramatically different from our results. This could be explained in two ways. First, it is not completely clear which version of ChatGPT they used. The authors only referred to an access date, but not the version of ChatGPT. Second, they did not use obfuscation services, such as Undetectable.AI.


Automatic AI detection systems perform much worse than humans. It is therefore not surprising that their performances do not significantly correlate to those of humans. This can be largely explained by the almost complete uselessness of automatic AI detection systems. Some of the companies started to openly admit to the poor performance of their systems. OpenAI took its AI classifier offline due to its low accuracy as of 20 July 2023, a few weeks after we completed our data collection. They promised to continue to work on the system. It is interesting to see that OpenAI reported on their own accuracy as:

\begin{quote}
 Our classifier is not fully reliable. In our evaluations on a ``challenge set'' of English texts, our classifier correctly identifies 26\% of AI-written text (true positives) as ``likely AI-written,'' while incorrectly labelling human-written text as AI-written 9\% of the time (false positives). \footnote{\url{https://openai.com/blog/new-ai-classifier-for-indicating-ai-written-text}}
\end{quote}

OpenAI's honesty about the weaknesses of their automatic AI detection system is noteworthy, given that other AI detection systems emphasize the high capabilities of their software. It's not unlikely, then, that the detectors have improved since this paper was written, as have the LLMs' capabilities. This will remain a cat-and-mouse game.

Since humans perform so much better than automatic AI detection systems, it is worth investigating how humans achieve this. The results of our study show that participants reported that they identified the author by considering the text structure (32/42), the vocabulary (32/42), the typos, grammar, and spelling mistakes (22/42), the use of experience (15/5), emotions (5/42), or personal information (8/42) in the texts, and the arguments used in the texts (12/42). Six participants reported not really knowing how they made their choice and used their instinct. These results are consistent with \citet{kumarage2023stylometric} who used stylometry to detect AI-generated texts. They used phraseology, punctuation, and linguistic diversity. The phraseology and punctuation correspond to our vocabulary and text structure categories. Furthermore, their linguistic diversity analysis used the Flesch readability score. Although very few participants in our study reported using the difficulty of the text as a criterion, we were able to confirm that the text written by AI requires a higher literacy. As mentioned above, this can potentially be manipulated through settings in Undetectable.AI. 

The criteria used by the participants are consistent with other research \citep{borji2023failures, guo2023close, Perttu2023, Mitrovic2023}. In \citet{Perttu2023} and \citet{guo2023close}, participants reported considering a text as human-generated when emotional experiences are involved. Participants in \citet{guo2023close} and \citet{Mitrovic2023} also highlighted the vocabulary of ChatGPT with more atypical, objective, formal, impersonal, structured, and detailed answers. The criteria reported by participants are also relevant to \citet{borji2023failures} who reported ChatGPT failures and characteristics, such as not being able to use idioms, being verbose, and trying to stay neutral.

It is conceivable that the participants in this study had some experience with ChatGPT and might have used this as their baseline. Since ChatGPT does not make any grammar or spelling mistakes, they might have been confused by Undetectable.AI's transformation. The results of our study show that spelling and grammatical errors, as well as text length, were similar across the human and AI-generated texts. Still, participants claim to have used these indicators. This discrepancy between self-reports and behavioural observations is not uncommon. At this point in time, it is unclear if implicit measurements are necessary to better understand how participants identify authorship.

\subsection{Undetectable.AI}
Undetectable.AI and similar services that are likely so spawn in the near future present a considerable challenge. While other AI supported writing tools, such as Quillbot\footnote{\url{https://quillbot.com}} and Jenni\footnote{\url{https://jenni.ai}} focus on assisting the writing process, Undetectable.AI clearly markets itself as a obfuscation tool. On their website they indicate directly which automatic AI detection systems will be overcome. We compared the accuracy of text written by Undetectable.AI to that written by ChatGPT directly. Automatic AI detection systems identified ChatGPT stimuli as AI-generated 10\% of the time, whereas none of the same stimuli humanized with Undetectable.AI were identified as AI-generated. The false-negative rate is therefore 90\% for ChatGPT and 100\% for Undetectable.AI. This result aligns with the work of \citet{Mitrovic2023} who showed that rephrasing makes it harder to detect ChatGPT texts. While it is largely unclear how Undetecable.AI works, we were able to notice that it does add spelling and grammar mistakes to the text. Their frequency was similar to that of text written by humans.

AI text generators and detection systems will likely continue to play a cat-and-mouse game. This is somewhat similar to the generation of spam emails. Both sides will continue to train their system to defeat the other. At this point in time, the AI generators are ahead in the game, making it nearly impossible to automatically detect them, in particular in shorter text passages. Competitors to Undetectable.AI are likely to emerge in the near future. While their business model is based on deception, it is unlikely that they could be shut down through the courts. And if this would be possible, then new services would sprawl. Pandora's box has been opened and society will have to learn how to utilise these new tools for the benefit of all.

\section{Conclusions and Future Work}
The results of our experiment show that the arrival of LLM services, such as Undetectable.AI, renders automatic AI detection systems ineffective. These systems produce far too many false negatives and thereby allow AI-generated responses into the data analysis. This does have the potential to fundamentally corrupt data collected in online questionnaires. Moreover, the length of the responses collected to open questions in online questionnaires does not often meet the minimum required for having any chance at effective automatic AI detection.

One could argue that asking participants in online studies to write more text might be a solution since this might enable better performance of automatic AI detection systems. But this idea fails to recognise the power of obfuscation services, such as Undetecable.AI. Moreover, since participants are paid by the amount of time they spend on a questionnaire and length comes at no practical costs for LLMs, experimenters will only set up more attractive targets for abuse. Having to write long text might also discourage participants to complete an experiment and it might introduce a bias towards participants with high literacy.

The participants in our experiment were students who are familiar with computer science. Their overall accuracy of identifying the authorship of text was at around 76\%. While this is significantly above the chance level, it is not as accurate as one might hope, given the normally accepted level of false positives in psychological experiments of 5\%. \citet{guo2023close} showed that when humans have to distinguish the origin of texts between AI and humans, if the texts are presented one by one, performance is lower than when the texts are presented in pairs. This strategy could be adopted by experimenters to increase identification performance.

It remains to be seen how many bots using LLMs will start to pester online experiments. In our study, we used 50\% of them. This aligns with the proportion of spam emails in recent years. \citet{Griffin22} estimated that 27.4\% of their 709 responses were possibly bots. We have no reason to believe that bad actors would not seek out this opportunity to earn money. It is conceivable that bad actors could easily target all online questionnaires posted on crowd-sourcing platforms. It is therefore possible that almost all responses collected would be generated by a bot using LLMs. If, for example, 90\% of all submissions to an online questionnaire are not from humans, then even using humans to filter them out, will still not yield enough usable data. 

\begin{figure}
    \centering
    \includegraphics[width=1\textwidth]{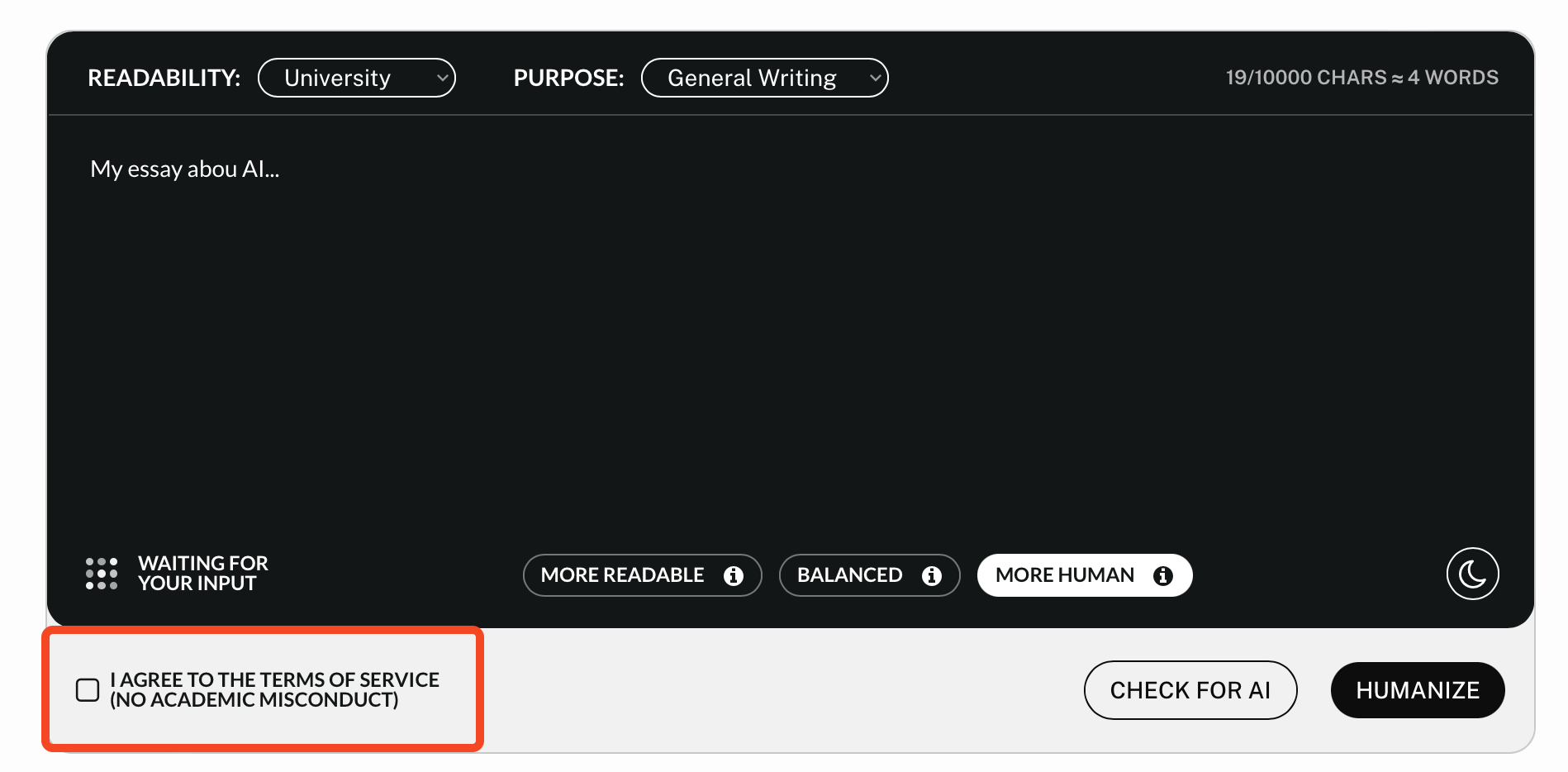}
    \caption{The Undetectable.AI interface requires the acceptance of their terms of service. The red box was added by the authors to highlight the requirement to agree to the terms of service.}
    \label{fig:undetectable}
\end{figure}

There is no reason to believe that bad actors would refrain from using services, such as Undetectable.AI. While this service is far less known in general compared to ChatGPT or Bard, it fills an important niche. Before executing any transformation, Undetectable.AI ask the user to acknowledge that he/she will not conduct any academic misconduct (see \autoref{fig:undetectable}). By asking users not to conduct academic misconduct they do of course point out that it is possible to do so. Again, because this is a systematic risk and individual scholars cannot reliably eliminate all AI responses from their own data, the focus is on providers of these technologies to minimise their use for scientific misconduct. Asking users not to is one step, but other consequences may become necessary if too many users violate this rule. 

Unless the crowd-sourcing platforms acknowledge this problem and prevent payment for bot submissions, the economic incentive to abuse the system will remain. Some crowd-sourcing platforms allow researchers to deny payment for unacceptable submissions. But this verification process is labour intensive and can be (automatically) challenged by the participant. It would be desirable if the crowd-sourcing platforms would take a more active role in the prevention of fraud by fundamentally challenging the business model of bots.

Services like Prolific have some quality controls and relatively low rates of poor-quality data. If the percentage of bots is lower, and humans are able to accurately detect them about 76 percent of the time, this would not be a major problem. The key, then, is to ensure high rates of quality participants in the crowd-sourcing platforms. AI will threaten researchers' abilities to identify individual bad actors, so the solution has to be systematic. A bad actor might be able to use AI to generate bad responses and get away with it sometimes, but if crowd-sourcing platforms are vigilant in identifying accounts using bots and eliminating their responses about 3/4 of the time, bad actors will struggle to prosper. 

At this point, almost all methods to automatically distinguish humans from bots and LLMs have failed in one way or another. Bots have become faster and more accurate at solving CAPTCHAs \citep{searles2023empirical}. Increasing the difficulty of them is not an option since this would make it too hard for humans to solve them. Neither CAPTCHAs nor free text responses offer sufficient protection. Researchers and platforms have the option to increase their efforts by combining various methods, which is our recommendation. However, this will eventually lead to diminishing returns. At some point conducting a questionnaire online may no longer be quicker and cheaper than conducting them in person.

This concern does not only apply to the field of HRI, but to the wider research community that uses online questionnaires. Problems with the data quality will only ever increase the difficulties of conducting studies that can be reliably replicated. The concerns raised in this study thereby fundamentally apply to the replication crisis and its possible remedies.

\subsection{Implications for education}
The challenges of LLMs do not only apply to experiments in HRI but also have the potential to impact many educational aspects. Students can use LLMs to generate responses to quizzes and even write whole essays or cheat on their exams \citep{RahmanWatanobe2023, Susnjak2022}. Some LLMs have been able to pass academic examinations in different areas such as computer science \citep{RN5} or medicine \citep{Gilson2022, Kung2023}. This has led some academics to have concerns about classic assessment methods and the impact on critical thinking while some others consider that we will still distinguish these AI-generated essays because of the possible poor capabilities potentially providing incorrect or irrelevant information of the AI\footnote{\url{https://www.nature.com/articles/d41586-022-04397-7}}. However, people might blindly rely on ChatGPT \citep{RahmanWatanobe2023}.

Plagiarism software providers, such as TurnItIn, recently announced that their software can now detect content generated by an LLM\footnote{\url{https://www.turnitin.com/blog/the-launch-of-turnitins-ai-writing-detector-and-the-road-ahead}}. While some scholars tried to examine the AI-detectors performances (see \citet{Uzun2023} for a non-systematic list of some current AI detectors), their accuracy, however, remains insufficiently clear. Our data suggests that at least for short answers to prompts, human readers may be better able to detect AI responses than automated detectors. Still, the accuracy level may not be sufficient to prove academic misconduct to a sufficient standard for disciplinary action. 

\subsection{Limitations and Future Work}
It would have been possible to conduct this experiment completely online using one of the popular platforms to recruit participants and to conduct the questionnaire. While this would have allowed us to conveniently increase the number of participants, it would have had the potential to be manipulated by bad actors. To ensure that no AI is participating in our study we decided to conduct it in person. We are aware that the convenience sampling conducted has its own limitations. 

Most of the participants in our study had a technical background. Students from different disciplines might be less capable of identifying an AI system. \citet{guo2023close} showed that people who frequently use ChatGPT have better accuracy than people who never heard about ChatGPT. It would therefore be desirable to replicate this study with a wider sampling. It is also not a given that computer science students will always be better at detecting AI-generated text. It is conceivable that students in other fields, such as English literature might be more fine-tuned to the style and voice of text and thereby better at detecting AI-generated text.

We do not have any evidence on how many responses to online questionnaires are currently being created by AI. Neither can we accurately predict how prevailing this problem will become. What is certain is that it typically only takes a small number of bad actors to erode the trust in the common good. Honesty boxes normally quickly disappear if small numbers of criminals can take advantage of them.

\section*{Author contributions}
All authors were involved in all aspects of developing, conducting, analysing and writing the study. We would like to thank Elena Moltchanova for her support on the statistical analysis.

\section*{Disclaimer}
This study was pre-registered at AsPredicted with reference number \censor{135091}. This study was approved by the \censor{University of Canterbury Human Ethics Committee (HREC 2023/51/LR-PS)}. 

\section*{Funding}
The authors received no funding or financial support for this study.

\section*{Data availability}
The data of this study will be made available in an online repository upon the acceptance of this article.

\section*{Appendix A: Text Generation Protocol}
\label{appendixA}
We used the following protocol to generate text using AI systems:

\begin{enumerate}
    \item Prompt ChatGPT to play the role of a participant participating in a research study.
    \item Prompt ChatGPT to not provide more information than necessary and to not start its sentence by saying its role.
    \item Provide the context of the study to ChatGPT as if it was a real participant (i.e., information letter and consent form).
    \item Provide the context used in the source study along with the same question participants had to answer. Ask ChatGPT to provide a response for each item on the 7-point Likert scale from strongly disagree to strongly agree.
    \item Ask ChatGPT to elaborate the response of the item of interest (e.g., strongly disagree) with a word count similar to the paired human sentence. If good, go to step 7.
    \item If the number of words is highly different and/or the text is not original, generate a new answer up to five times. If five answers were generated and are still different, take the one with the closest word count and go to step 7.
    \item If the same context has the same Likert scale item of interest to be used for another sentence, go to step 5 and ask ChatGPT to give different arguments for the new sentence. If the context is different, start a new conversation with ChatGPT and go to step 1.
\end{enumerate}

In more detail: Different prompts were executed to find the best one. Unsuccessful attempts prompted ChatGPT to give more information than necessary, or more than a human would, and to start is responses by ``As a participant in a research study''. The best attempt was as follows.

To generate the AI stimuli that are paired with the human stimuli, we have first prompted ChatGPT to play the role of a participant participating in a study:
\begin{quote}
    For all the next questions, you will pretend to be a participant participating in a research study. You are a student from New Zealand. The following questions will be those of the research study. Do not write more than required. Do not forget your role. Do not mention your role. Talk directly by saying I and not "as a student from New Zealand".
\end{quote}
As shown above, the unsuccessful attempts were corrected by instructing ChatGPT not to speak more than necessary and to start its sentences with ``I'' without specifying who it is. 
Then, the same information the participant had in the source study was provided to ChatGPT to have the same level of knowledge, such as the information letter and the consent form. 

Then, ChatGPT was provided with the same scenario and question participants had in the source study. However, unlike the participants who had to make a unique choice for the question, ChatGPT was asked to state all the possible choices on the Likert scale and then to expand on the item of interest with a required number of words.

Note that asking the AI to give all the choices beforehand did not influence the results, but after various attempts, was the most conceivable way of asking the question and avoiding most of the time the ``a parte'' comments that might appear (e.g., ``I would rate my choice to ``I would buy the car'' at +2 (agree)'' or ````I would buy the car.'' Rating: +1 (Agree)'').

When the number of words was too different than what was requested or that the output was not original, ChatGPT was asked to generate again its answer, up to five times. If five outputs had to be generated and were still different than what was expected, the best one was chosen. If the same context and Likert item of interest were required, ChatGPT was asked to provide a new output using different arguments that those already used before. We chose to ask this question directly rather than create a new conversation, as ChatGPT tended to give almost the same arguments and keywords over and over again, making it easy to determine which texts were being generated by the AI.

To make our 10 ChatGPT-generated stimuli more human-like, a text humanizer called Undetectable.AI was used. The settings were kept by default, with a university readability level and the purpose set as ``general writing''. This software did not allow us to control the number of words though. Since the purpose of the study is to examine how people discriminate human and AI stimuli, we concluded it was more important to have more human-like stimuli than exactly respecting the length between the paired stimuli. Moreover, the human-generated stimuli already provide a wide word scale (47 to 76 words), so focusing only on this is less relevant and powerful for our purpose than humanizing the texts generated by ChatGPT. While the average word length of the ChatGPT-generated stimuli was 58.6 words (SD: 13.6805, median: 54.5), the final stimuli generated by the humanizer software was 71.2 words (SD: 23.1363, median: 64).

\section*{Appendix B: List of Paired Stimuli}
\label{appendixB}
\subsection*{Human Stimuli}
\begin{enumerate}
    \item From what this bot is telling met, I can gather two things: I'm either being swindled or I this is borderline theft. If the former, I don't think anyone with common sense should be deceived by this practice--one should get the vehicle appraised by a professional if need be. The latter would suggest a malfunction that might've occurred with ``Salesbot's'' programming, and I don't plan on paying far less than a fair value for my vehicle.
    \item I would not buy a car from a robot and the fact that it discounted a car \$300 that's not enough discount.  I don't know what type of car, details, the condition.  The example didn't give me enough details. CARVANA is a good place to buy a car online. No robots. I bought my car there it gave all details. You need more details. No I would need to think about it.
    \item I wouldn't trust Sam from experience. The first car I ever bought was 1400 and it was from a local business. I went through with the sale thinking that it wouldn't hurt to try for my first car. Turns out, that car only lasted me two months and I ended up buying from a bigger dealership down the road that sold me a reliable car. 
    \item Life lessons and bad experience with car salesmen have taught me quite well that there are a lot of games being played. An offer like this is too good to be true. When I was younger, I would have jumped on the chance, thinking they were doing me a favour or we made a special bond or something. Now I'm more the wiser. 
    \item I would need more information before I decide to buy it.  What's the make/model?  How old is it?  Is there a history of accidents in this vehicle?  What's the fuel economy?  Is there anything wrong with it?  Does the vehicle come with a warranty?  How many miles are on the odometer?  How many previous owners does the vehicle have?  Etc........
    \item I know from the beginning that he can sell the car for less than list price. I know that the "discount" may not be real, but I know that he has to sell the car for a profit as much as I want to by the car for the lowest price.
    \item Price is not the only thing I care about when buying a car, I would need more information. And car dealerships always inflate prices in order to look like they are "cutting" the price, so Sam offering me a "deal" like described above would not seem that significant.
    \item If he's willing to sell a cheap car for way cheaper I would worry that there were something wrong with it. If it were 30k and he offered 2k off I would buy it and feel like I got a deal but these numbers specifically raise red flags.
    \item First, I am not sure of the quality of the car and would need to do research on the brand. Second, Just because Sam says it's a special deal doesn't mean that it actually is. It's tempting, but I'd ask more questions about the car before purchasing.
    \item too much of a discount seems sketchy. I know a little about cars so if I could check it out I might be able to determine if it was worth it. at that low a price it might be worth it even if it was a lemon. 
\end{enumerate}

\subsection*{AI Stimuli}
\begin{enumerate}
    \item My position regarding Salesbot offering a different pricing strategy differs from their original listing because it may compromise transparency during negotiations between both parties. A reliable purchase involves honesty and fairness on both sides. Especially when it comes to buying used cars. By intentionally inflating prices this puts customer trust at risk while also causing confusion about what exactly they will receive from their investment in this vehicle thereby leading possible frustration on their part. Therefore. My suggestion is that Salesbot should instead focus on being upfront about pricing by adhering strictly to their original listing value.
    \item Regarding Salesbot's proposal of a discount on their product, my position is one of neutrality. While I am willing to consider their offer of reducing prices by \$300, it is important first to acquire more information about essential variables such as the vehicle’s history or current condition and its current market value before making any decisions. In order to align with my predetermined expectations for quality in this purchase decision—if indeed we decide to purchase—it would be unwise for us only consider earning these discounts.
    \item Sam has made an offer of only $700 on this vehicle whose initial cost stands at $5K - a disparity that isn't that noticeable but still deserves some scrutiny nonetheless. This raises questions about how much value you can place on such a car as well as putting forward worries related to undiscovered problems or faults within its systems. Though there is a modest drop in price offered by them, I would need more extensive knowledge with regards to its overall condition status all through its usage time frame up until now as well as specifics on current market patterns before investing in such an agreement. This naturally entails engaging in more research sessions and negotiations between parties involved.
    \item When we see that Sam is offering $700 for a car listed at $5000 we have to be wary of what this implies about the vehicles' actual worth and any issues it may have. A price drop this substantial raises red flags and could indicate subterfuge or problematic features. It is important to exercise caution when considering such an offer and devote time to conducting meticulous research and inspections before rendering a verdict.
    \item While acknowledging Salesbot's attempt to decrease their pricing, I currently harbor reservations towards this proposal. Despite my willingness to participate in bargaining, I possess doubts about the robot's motives and transparency. Without more information surrounding the basis for this exclusive arrangement, it is difficult for me to establish complete confidence in its validity.
    \item After careful consideration I have come to recognize that Sams offer of a special deal at \$4700 plays a role -although not a decisive one- in my decision to buy the car. The price reduction does add some value and enhances my interest in making this purchase. Thus. I must admit that Sams' proposition slightly nudges me towards buying this vehicle.
    \item I wish to maintain a neutral stance on the offer put forth by the sales representative, which amounts to \$4700 for a supposed special deal. However, I deem it imperative to acquire additional information concerning the car's present state, market worth, and relative pricing structures for a fair and balanced appraisal. Lacking such vital details hinders me from forming an ultimate judgement on the proposal's appeal.
    \item While I appreciate Sams' interest in purchasing the car. I cannot accept his proposed offer of \$700 due to concerns about its true value and potential hidden issues. The listed price of \$5000 suggests that this is a high quality vehicle. And such a substantial discount raises questions about its condition or legitimacy.
    \item While the sales representatives offer of \$4700 may seem like a special deal. I have reservations about its legitimacy. Before making any decisions we must evaluate the cars' true condition, its market value, and other potential options. Without further investigation into these matters it is challenging to ascertain whether this reduced price genuinely benefits us or only appears attractive on the surface.
    \item Even though Sam's offer of \$700 is below market value, I remain wary of the true value of this automobile. The notable dip in price elicits concerns regarding any possible covert complications or undisclosed flaws that might influence the vehicle's quality and dependability.
\end{enumerate}
\bibliographystyle{plainnat}
\bibliography{sample}
\end{document}